\documentclass[11pt,a4]{article}

\usepackage{geometry}
\usepackage{amsmath}    % need for subequations
\usepackage{graphicx}
\usepackage{natbib}
\usepackage{multirow}
\usepackage{verbatim}
\usepackage{setspace}
\usepackage{float}

\usepackage{amssymb}
\usepackage{subcaption}
\usepackage{latexsym}
\usepackage{color}
\usepackage{comment}

\usepackage{graphicx}
\usepackage{caption}

\usepackage{float}

\restylefloat{table}

\usepackage{epstopdf}
\usepackage{comment}
\usepackage{bm}
\usepackage{eucal}
\usepackage{mathrsfs}
\usepackage{graphicx}
\usepackage{booktabs}
\usepackage{color}

\DeclareMathAlphabet{\mathpzc}{OT1}{pzc}{m}{it}
\DeclareMathAlphabet{\mathcalligra}{T1}{calligra}{m}{n}
\include{GH}

\newtheorem{proposition}{Proposition}
\title{\Large \bf Predictive Inference for Spatio-temporal Precipitation Data and Its Extremes\footnote{The authors would like to thank Prof. Noel Cressie who provided insightful comments on this paper.}
}
\date{Version of October 17, 2014}
\author{{\bf Yang Liu\footnote{Corresponding Author. Email: Yang.Liu@csiro.au} $\>$and Philip Kokic}\\
CSIRO, Canberra, Australia}

\begin{document}
\newgeometry{left=1.85cm,right=1.85cm, bottom=2.25cm,top=1.75cm}
\maketitle
\centerline{\footnotesize Under review at the {\it Journal of the American Statistical Association}}
\begin{abstract}
\footnotesize
Modelling of precipitation and its extremes is important for urban and agriculture planning purposes. We present a method for producing spatial predictions and measures of uncertainty for spatio-temporal data that is heavy-tailed and subject to substaintial skewness which often arise in measurements of many environmental processes, and we apply the method to precipitation data in south-west Western Australia. A generalised hyperbolic Bayesian hierarchical model is constructed for the intensity, frequency and duration of daily precipitation, including the extremes. Unlike models based on extreme value theory, which only model maxima of finite-sized blocks or exceedances above a large threshold, the proposed model uses all the data available efficiently, and hence not only fits the extremes but also models the entire rainfall distribution. It captures spatial and temporal clustering, as well as spatially and temporally varying volatility and skewness. The model assumes that the regional precipitation is driven by a latent process characterised by geographical and climatological covariates. Effects not fully described by the covariates are captured by spatial and temporal structure in the hierarchies. Inference is provided by MCMC using a Metropolis-Hastings algorithm and spatial interpolation method, which provide a natural approach for estimating uncertainty. Similarly, both spatial and temporal predictions with uncertainty can be produced with the model.
\end{abstract}
\section{Introduction}
\begin{comment}
Relatively short, intense bursts of rainfall, such as severe thunderstorms can cause flash flooding. It can occur in most parts of the world, especially in tropical regions and poses the greatest threat of loss of life. In other cases, prolonged rainfall sometimes produces runoff which overflows the banks of the watercourse. These floods can also result in significant property damage and major social disruption. \\
\end{comment}
Statistical modelling of precipitation has important applications in many different fields of research including hydrology, agriculture, and environmental sciences. Within each of these fields information is required on a number of spatial and temporal scales. In hydrological applications the frequency and duration of extreme precipitation events over short time periods is very important \citep{Shao2013, Fowler2007,Tetzlaff2005}. For agricultural modelling climate information is required on the short-term variations associated with extreme and non-extreme events through the realistic simulation of observed data \citep{Keating2003, Kokic2013}, as well as intermediate variations associate with seasonal and inter-seasonal variations, and long-term variations due to causes such as climate change. Environmental science applications often require precipitation data with a high degree of spatial resolution \citep{Ashcroft2011,Ashcroft2012}. For these reasons, statistical approaches that can provide consistent information covering these varying temporal and spatial characteristics is valuable. To address these requirements we need a flexible statistical approach that can accurately represent extreme events, and varying shapes and skewness of the rainfall distribution, as well as reliable estimates of the serial and spatial dependencies in these data. \\
\indent The objective of this paper is to describe a unified statistical model that moves towards meeting several of these objectives through the use of a Bayesian hierarchical model that utilises generalised hyperbolic (GH) processes. This method has significant advantages over models built upon statistical extreme value theory (EVT). The proposed model allows one to study both high and low precipitation events in a unified model. It is more data efficient, because it does not involve modelling maxima of finite-sized blocks or choosing a large threshold, so a much larger sample that may contain additional information will contribute to the estimates. 

\indent This research focuses on hydrometeorological data, but we note that our spatio-temporal model is not limited to this context, and this methodology can be adapted to other types of data and applications.

\subsection{Measures for precipitation and extremes}
Estimates of potential flooding and rainfall deficiency are necessary for city, rural area development planning and risk assessment. A commonly used measure of extreme events is the return period. It is a statistical measurement typically based on historic data denoting the average recurrence interval over an extended period of time. The calculation of return period assumes that the probability of the event occurring does not vary over time and is independent of past events. Take $p=1/T=1-P(X\le x)$, where $T$ is the return interval, $P(X\le x)$ is the cumulative distribution function of rainfall and $x$ is an extreme event. Practically, for $x$ large enough, the independence of events can be assumed. If the probability of an event occurring is $p$, then the probability of the event not occurring is $q=1-p$. The binomial distribution can be used to find the probability of occurrence of an event $r$ times in a period of $n$ years:
\begin{align}
\begin{pmatrix}
  n  \\
  r 
 \end{pmatrix}\times p^r\times q^{n-r}.
\end{align}
\indent Other frequently used uncertainty measures for precipitation and the extremes are duration over/below thresholds, as well as duration of zero precipitation. For aggregated data (eg. monthly and yearly), the number of time units over thresholds are more appropriate measures. These quantities provide probabilistic measures for prolonged extreme events and are highly appreciated in hydrological and environmental research.

\subsection{Extreme value statistics}
Rainfall arises from physical processes, but it is widely known that physical models such as General Circulation Models (GCMs) are inadequate for extremes, due to their coarse spatial resolution and the current incomplete understanding of the climate system \citep{ye2011method}. Statistical models are therefore often considered when modelling extreme precipitation events.\\
\indent Extreme value theory (EVT) is a frequently used approach for modelling extremes, because it provides statistical models for the tail of a probability distribution and complements to modeling the mean or central part of a distribution. EVT is based on the asymptotic arguments that lead to the generalized extreme value (GEV) distribution. For simplicity, in the univariate case, given independent and indentically distributed continuous data $Z_1, Z_2,\dots, Z_n$, and letting $M_n=\max\left(Z_1, Z_2, \dots, Z_n\right)$, it is known that if the normalized distribution of $M_n$ converges as $n\to\infty$, then it converges to a GEV distribution \citep{fisher1928limiting,gnedenko1943distribution,von1936distribution}. Because of its asymptotic justification, the GEV distribution is used to model maxima of finite-sized blocks such as annual maxima. For example, this method was used by \cite{nadarajah2007maximum} and \cite{feng2007modeling} for modelling annual rainfall maxima in China and South Korea respectively. The model was fitted to individual weather stations, and the spatial distribution of the extreme rainfall return period was calculated through spatial interpolation. A more sophisticated model was used by \cite{gaetan2007hierarchical}. The authors analysed annual rainfall maxima at weather stations in northeastern Italy using nonstationary spatial dependence and a random temporal trend in the parameters of the GEV distribution.\\
\indent However, when daily precipitation is available, models which only use each year's annual maximum disregard other extreme events. The Generalized Pareto (GP) distribution is based on the exceedances above a threshold \citep{pickands1975statistical}. Exceedances (the amounts which observations exceed a threshold $u$) should approximately follow a GP distribution as $u$ gets large and the sample size increases. In this case, the tail of the distribution is characterized by the equation
\begin{align}
P\left(Z>z+u\vert Z>u\right)=\max\left(0,\left(1+\kappa\frac{z}{\sigma}\right)^{-1/\kappa}\right).
\end{align}
\indent The scale parameter $\sigma$ must be greater than zero, and the shape parameter $\kappa$ controls whether the tail is bounded $(\kappa<0)$, light $(\kappa\to0)$, or heavy $(\kappa>0)$. The GP approach has been widely applied in the literature. \cite{cooley2007bayesian} used the method with a common threshold at all weather stations to map return levels for extreme precipitation in Colorado. A stationary isotropic exponential covariance function was used to induce spatial dependence for these parameters. The shape parameter $\kappa$ had two values depending on the station's location. \cite{turkman2010asymptotic} contructed a similar but more complex model for space-time properties of wildfires in Portugal, using a random walk to describe the temporal properties, and smoothing for the spatial modelling of exceedances. \cite{ex2} and \cite{ex1} also used GP spatial models for similar environmental data.\\
\indent In practice, a threshold is chosen at a level where the data above it approximately follows a GP distribution and the shape and scale parameters are estimated. In order for the exceedances to follow a GP distribution, the chosen threshold is often large, so an enormous amount of data that could potentially provide additional information is discarded. In addition, threshold selection can be rather subjective. It is usually done using diagnostic plots that show how quantities such as the shape parameter vary as the threshold changes. Once chosen, the uncertainty associated
with the choice of threshold is not accounted for \citep{coles1996bayesian}. Furthermore, \cite{cooley2007bayesian} argued that the low-precision of rainfall data can also introduce a problem with threshold selection.\\
\indent Ideally we would like to develop a more flexible spatio-temporal model which overcomes these problems. To this end, we need a class of distributions which can approximate the power law decay of the GP distribution in the tails, as well as provide flexibility in the centre and shoulder of the distribution. To meet these requirements we consider the generalised hyperbolic (GH) distribution.
\subsection{The generalised hyperbolic distribution}
The GH distribution was first introduced by \cite{barndorff1977} in connection with dune movements modelling. The Lebesgue density of the GH distribution is defined as 
\begin{align}
\label{newformulation}
d_{GH(\lambda,\chi,\psi,\boldsymbol\mu,\boldsymbol\Sigma,\boldsymbol\gamma)}(\boldsymbol x)&=\frac{\left(\sqrt{\frac{\psi}{\chi}}\right)^\lambda\left(\psi+\boldsymbol\gamma^2 \boldsymbol\Sigma^{-1}\right)^{\frac{1}{2}-\lambda}}{\sqrt{2\pi\left\vert\boldsymbol\Sigma\right\vert}K_\lambda\left(\sqrt{\chi\psi}\right)}\notag\\
&\frac{K_{\lambda-\frac{1}{2}}\left(\sqrt{\left(\chi+\left(\boldsymbol x-\boldsymbol\mu\right)^2\boldsymbol\Sigma^{-1}\right)\left(\psi+\boldsymbol\gamma^2\boldsymbol\Sigma^{-1}\right)}\right)\exp{\left(\left(\boldsymbol x-\boldsymbol\mu\right)\Sigma^{-1}\boldsymbol\gamma\right)}}{\left(\sqrt{\left(\chi+\left(\boldsymbol x-\boldsymbol\mu\right)^2\boldsymbol\Sigma^{-1}\right)\left(\psi+\boldsymbol\gamma^2\boldsymbol\Sigma^{-1}\right)}\right)^{\frac{1}{2}-\lambda}},
\end{align}
where $K_v$ is the modified Bessel function of the third kind with index $v$, $\lambda\in\mathbb{R}$, $\chi, \psi\in\mathbb{R}^{+}$, $\boldsymbol\mu, \boldsymbol\gamma\in\mathbb{R}^N$ and $\boldsymbol\Sigma\in\mathbb{R}^{N\times N}$. To gain some intuition of the expression above, one often writes the generalised hyperbolic distribution as the following mean-variance mixture. A random variable $\boldsymbol X$ is said to have a GH distribution if
\begin{align}
\label{mvm}
\boldsymbol X=\boldsymbol\mu+\boldsymbol\gamma W+\sqrt{\boldsymbol\Sigma}\sqrt{W}\boldsymbol Z,
\end{align}
where $\boldsymbol Z$ is a $N$-dimensional normal random variable with zero mean and unit variance, and $W$ has a generalised inverse Gaussian distribution with parameters $\lambda, \chi$ and $\psi$, i.e. $W\sim GIG(\lambda, \chi, \psi)$. Since $\boldsymbol X$ given $W=w$ is Normal with conditional mean $\boldsymbol\mu+\boldsymbol\gamma w$ and variance $\boldsymbol\Sigma w$, it is clear that $\boldsymbol\mu$ and $\boldsymbol\Sigma$ are location and dispersion parameters, respectively. There is a further scale parameter $\chi$, a skewness parameter $\psi$ to allow for flexible tail modelling; and the scalar $\lambda$, which characterises certain subclasses, also influences the size of mass contained in the tails \citep{barndorff1977}.\\
\indent One of the appealing properties of normal mixtures is that the moment generating function of a GH random variable $\boldsymbol X$ can be easily calculated using the moment generating function of the generalised inverse Gaussian distribution. In particular, the mean and variance are given by
\begin{align}
\label{mv}
E(\boldsymbol X)=\boldsymbol\mu+\boldsymbol\gamma E(W)\>\>\>{\text{\normalfont and}}\>\>\>Var(\boldsymbol X)=\boldsymbol\gamma^2 Var(W)+\boldsymbol\Sigma E(W).
\end{align}
\indent The GH distribution, as the name suggests, is of a very general form, and contains as special cases many important distribution widely used in applications. It includes, among others, Student's t-distribution, the Laplace distribution, the hyperbolic distribution, the normal-inverse Gaussian distribution and the variance-gamma distribution. It is often used in economics, with particular application in the fields of modelling financial markets and risk management \citep{eberlein2004generalized}, due to its flexiblity and heavy tails.\\
\indent There are several parameterisations of the GH distribution. The $(\lambda,\chi,\psi,\boldsymbol\mu,\boldsymbol\Sigma,\boldsymbol\gamma)$ parameterisation used in this paper has the drawback of an identification problem \citep{barndorff2001normal}, i.e. $GH(\lambda,\chi,\psi,\boldsymbol\mu,\boldsymbol\Sigma,\boldsymbol\gamma)$ and $GH(\lambda,\chi/k,k\psi,\boldsymbol\mu,k\boldsymbol\Sigma,k\boldsymbol\gamma)$ are identical for any $k>0$. This is because that $\chi$ and $\left\vert\boldsymbol\Sigma\right\vert$ are not separately identified. This problem can be solved by introducing a suitable constraint on the parameters. \cite{barndorff2001normal} fixed the determinant of $\boldsymbol\Sigma$ to be 1. However under this setup, it is difficult to standardise the GH distribution so that it has mean zero and unit variance. Such standardisation is necessary when developing our hierarchical model. Fortunately, the identification problem can also be solved by fixing $\chi=1$, and that leads to a simple standardisation method \citep{JEd}, which we restate here in the following Proposition. 
\begin{proposition}
\label{prop1}
Let us modify \eqref{newformulation} by replacing $\boldsymbol\gamma=\boldsymbol\Sigma\boldsymbol\tau$, $\boldsymbol\tau>0$, we have $\boldsymbol X\sim GH(\lambda,\chi,\psi,\boldsymbol\mu,\boldsymbol\Sigma,\boldsymbol\tau)$. For any $\lambda, \psi, \boldsymbol\tau$ and fixed $\chi=1$, if $\boldsymbol\mu$ and $\boldsymbol\Sigma$ satisfy
\begin{align}
\label{muvar}
\boldsymbol\mu=-L\left(\boldsymbol\tau,\psi,\lambda\right)\boldsymbol\tau\>\>\>{\text{\normalfont and}}\>\>\>\boldsymbol\Sigma=\frac{\sqrt{\psi}}{M_{\lambda+1}\left(\sqrt{\psi}\right)}\left(\boldsymbol I_N+\frac{L\left(\boldsymbol\tau,\psi,\lambda\right)-1}{\boldsymbol\tau'\boldsymbol\tau}\boldsymbol\tau\boldsymbol\tau'\right),
\end{align}
where
\begin{align}
\label{s1}
&M_{\lambda+1}\left(\sqrt{\psi}\right)=\frac{K_{\lambda+1}\left(\sqrt{\psi}\right)}{K_{\lambda}\left(\sqrt{\psi}\right)},\>\>N_{\lambda+2}\left(\sqrt{\psi}\right)=\frac{K_{\lambda+2}\left(\sqrt{\psi}\right)K_{\lambda}\left(\sqrt{\psi}\right)}{K_{\lambda+1}\left(\sqrt{\psi}\right)^2}\>\>\>{\text{\normalfont and}}\\
\label{s2}
&L\left(\boldsymbol\tau,\psi,\lambda\right)=\frac{\sqrt{1+4\left(N_{\lambda+2}\left(\sqrt{\psi}\right)-1\right)\boldsymbol\tau'\boldsymbol\tau}-1}{2\boldsymbol\tau'\boldsymbol\tau \left(N_{\lambda+2}\left(\sqrt{\psi}\right)-1\right)},
\end{align}
then $E(\boldsymbol X)=0$ and $Var(\boldsymbol X)=1$.
\end{proposition}
An asymptotic justification for the use of the GH distribution can be found in \cite{mypaper0}. The authors showed that for any chosen threshold, the GH distribution can approximate the power law decay of the GP distribution in the tails, if the given GP distribution has $\kappa>0$ or $\kappa\to0$. They also demonstrated that the GH model can achieve a good fit at the centre and shoulders of the distribution by applying the model to various rainfall datasets. In addition, fitting a GH model does not involve choosing an appropriate threshold which can be a major drawback of GP models because they do not account for the uncertainty associated with the choice of threshold, and potentially useful information below the threshold is discarded for estimating the extremes. This is a significant advantage of modelling with the GH distribution as it allows us to make inferences of not only the rainfall extremes, but also the entire distribution. Consequently the prediction credible intervals (CI) for return periods of extreme events and other statistical inference obtained under the GH model have a larger sample contributing to the estimates.

\subsection{Paper outline}
The next section describes the precipitation data sources and some preliminary analysis. In Section 3 we describe the structure of our Bayesian hierarchical model. We discuss the GH-based hierarchical model for rainfall and rainfall extremes in Section 3.1, discuss the Bayesian framework including prior distributions in Section 3.2, briefly describe our MCMC method for model inference in Section 3.3, and discuss how our model was used to produce spatial predictions in Section 3.4, and briefly illustrate temporal prediction in Section 3.5. We then present results, including spatial predictions and model validation in Section 4. Finally, we conclude with a discussion in Section 5.

\section{Data}
Winter rainfall in south-west Western Australia (SWWA) was once considered the most consistent and reliable anywhere in Australia. However, around the mid 1970s, there was a shift to more volatile winter conditions, which has continued to this day, so the region can experience both extreme and sometimes insufficient rainfall in winter \citep{HE}. These conditions have had strong negative impacts on urban infrastructure, surface and ground water supplies, agriculture and natural ecosystems.\\
\indent Although these events are rare, understanding the frequency, intensity and duration is important for public safety and long-term planning. In this paper we illustrate the proposed method by applying it to precipitation data from 21 weather stations in the Natural Resource Management (NRM) Region 504 of SWWA; see Figure 1. Our method aims to produce out-of-sample spatial predictions of precipitation return periods, duration over/below thresholds and relevant prediction maps. 
\subsection{Study region, weather stations, and covariates}
Daily precipitation (mm) data are available from 21 meteorological monitoring sites in the study region obtained from the Australian Bureau of Meteorology website (http://www.bom.gov.au); see Figure 1. In this paper, we analyse daily precipitation data for 92 days in the months of June-August for 48 years from 1960 to 2008. Hence, a total of 92,736 data points for daily precipitation are used in this analysis, of which 1.10\% were missing, and 38.27\% were recorded 0 mm. \\
\begin{figure}[h]
 \centerline{\includegraphics[trim = 0mm 6.5mm 0mm 29mm, clip,width=110mm]{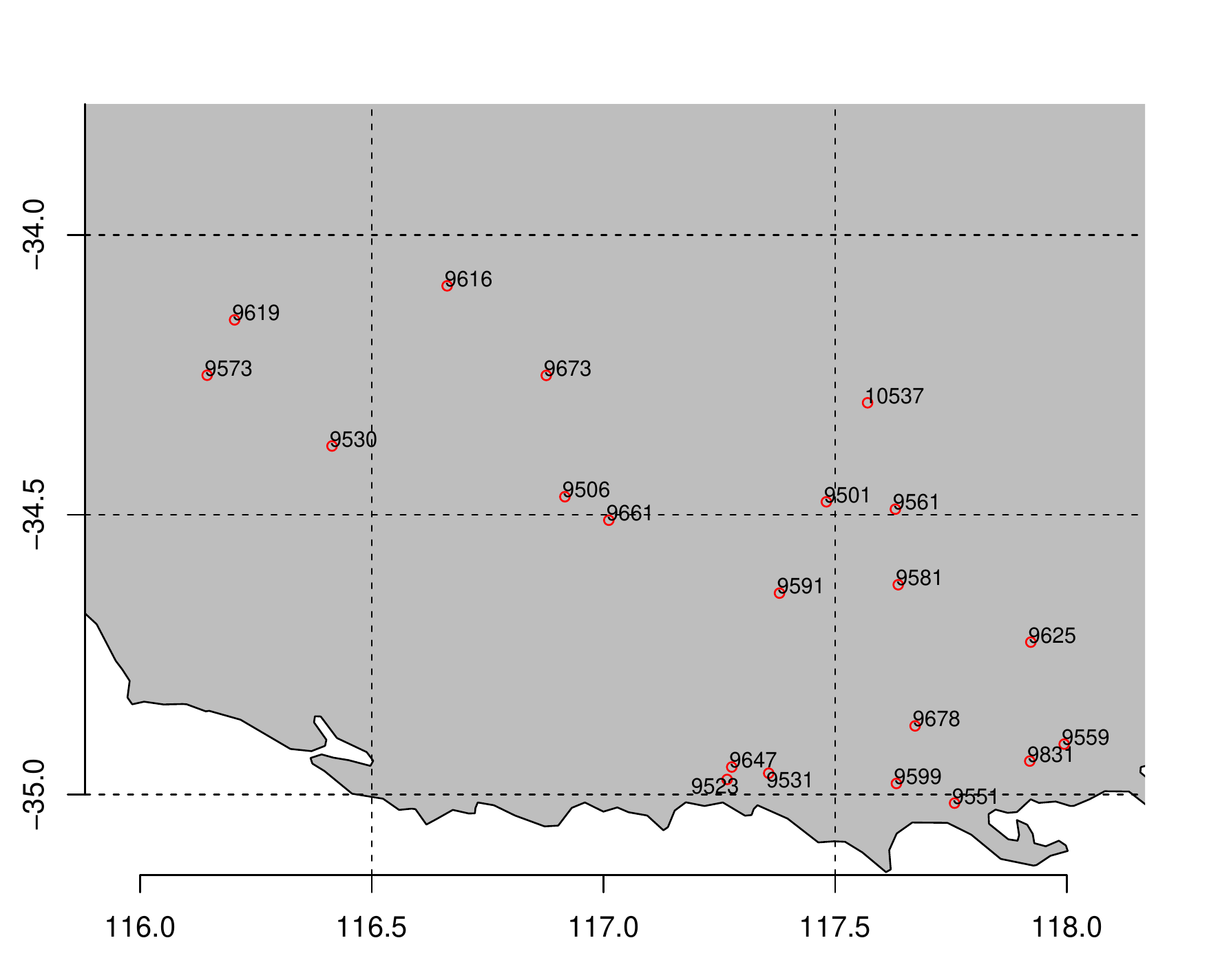}}
 \caption{Weather stations in the the south-western agricultural zone of Western Australia (NRM Region 504). Location numbers are given next to the circles.}
\end{figure}
\indent  This paper also accounts for various climatic drivers that have an influence on precipitation levels. Understanding the effects of climatic drives on rainfall and the pattern of seasonal precipitation is important for assessing agricultural productivity and
water risk-management. Empirical evidences on the relationships between them are discussed in many studies, e.g. \cite{schepen2012evidence}. Three important climatic drivers: the El Ni{\~n}o southern oscillation anomaly (NINO 3.4), Southern Hemisphere Annular Mode (SAMI) and the Indian Ocean Dipole (IOD) are considered in this study to understand their effects on precipitation levels. It has been observed that these indices have potential teleconnection with precipitation levels. The NINO 3.4 index is a monthly time series of mean sea surface temperature (SST) from an equatorial region of the eastern Pacific that covers 5$^{\circ}$ south to 5$^{\circ}$ north and 125$^{\circ}$ west to 175$^{\circ}$ east \citep{schepen2012evidence}. The SAMI index includes the Antarctic oscillation and is obtained by the differences in the normalised monthly zonal mean sea level pressure between 40$^{\circ}$ and 70$^{\circ}$ south \citep{nan2003relationship}. The IOD index is calculated from monthly SST anomalies in the equatorial Indian Ocean \citep{saji1999dipole}. Another important variable, elevation (ELE), has also been considered in our model as it has a strong correlation with rainfall in many places.\\
\indent We have not accounted for any seasonal effects in our data. Restricting our analysis to the winter months reduces seasonality and on inspecting the data from several sites there was no obvious seasonal effect. Likewise, we have not accounted for any temporal trends in the data. However, our series are relatively short and it may be difficult to discover any trend in winter precipitation over the last 45 years. Furthermore, the purpose of this study is not focused on climate change. Both climate change and seasonal effects would be simple extensions of this study. For example, the model proposed in this paper can be generalised to a dynamic linear model \citep{west1985dynamic,stroud2001dynamic}, which is popular for modelling data with seasonal variations. Recent applications include \cite{dou2010modeling,ghosh2010spatio, mahmoudian2014spatio} and \cite{sahu2012comparison}.
\subsection{Exploratory analysis}
Table 1 provides summary statistics for the variables used in the model. Daily rainfall ranges from 0 mm to 112.00 mm with a mean value of 3.73 mm. The table also provides summary statistics for the explanatory variable. NINO 3.4 ranges from 24.43 to 29.14, the SAMI ranges from -7.13 to 5.36 and the IOD varies from -2.74 to 3.55 units for the 45-year period considered in this study. Finally the highest elevation of 300 metre is at location 3 in Figure 1.\\
\begin{table}[h]
\footnotesize
\centering
\begin{tabular}{llll}
\toprule 
\toprule 
  Variables  & Minimum & Mean &  Maximum  \\ 
    \midrule
Rainfall (mm)&0.00 & 3.73& 112.00 \\ 
NINO 3.4 &24.43 &27.01& 29.14 \\ 
SAMI &-7.13&0.14& 5.36  \\ 
IOD & -2.74 & 0.01&3.55 \\ 
ELE (metre)&10.00&185.40& 300.00  \\ 
    \bottomrule
\end{tabular}
 \label{sumt} % is used to refer this table in the text 
\caption{\footnotesize Summary statistics for the variables used.} % title of Table 
\end{table} 
\indent The wettest year was 1987 with an average winter daily rainfall of 4.18 mm, while the dryest was 1974, with an average winter daily rainfall of 1.12 mm. The largest observed number of days with zero rainfall (33 days) occurred in 1978 and the smallest (17 days) in 1986 and 2005. The data shows temporal dependencies and non-stationarity. Also the level of rainfall differs from site to site which may lead to heavier rainfall or drier conditions in some areas.

\section{Model specification}
In this section we describe the general structure of the proposed hierarchical model in detail. Hierarchical models have been widely used in spatio-temporal modelling and they allow one to statistically model a complex process and its relationship to observations in several simple components. For an introduction to such models, see \cite{gelman2013bayesian}.

\subsection{General structure of the hierarchical model}
Let $D_t(s_i)$ denote the observed point referenced data at location $s_i$, $i=1,\dots,N$, and time $t$, $t=1,\dots,N$. We are interested in making inference on the $D$ process on the basis of data. There are three layers in our hierarchical model. In the first level of modelling we write:
\begin{align}
D_t(s_i)=Y_{t}(s_i)I_{\{Y_{t}(s_i)\in E\}},
\end{align}
where $I_{\{\cdot\}}$ is the standard indicator function, and the set $E=(\infty,0]$ contains values for which $Y_t(s_i)$ is observed. Similiar censoring approach for modelling zero precipitation was also used in \cite{12qwe}, \cite{glasbey1997rainfall} and \cite{ex1}. We will discuss alternative methods in Section 5. The mean process is modelled in the second level of the hierarchy:
\begin{align}
Y_{t}(s_i)=\beta_0+\sum^M_{k=1} \beta_k X_{kt}(s_i)+ V_t(s_i),
\end{align}
where $\beta_0$ represents the global intercept, $X_{kt}(s_i)$ is the $k$th regressor with regression coefficient $\beta_k$. In the final stage of modelling we have the process for the spatially and temporally correlated error $V_t(s_i)$:
\begin{align}
&V_{t}(s_i)=bV_{t-1}(s_{i})+U_t(s_i),\\
&U_{t}(s_i)=\sum^N_{j=1}\sigma_{t}(s_i,s_j)Z_{t}(s_j),\\
&\sigma_t^2\left(s_i,s_i\right)=\omega+\alpha U^2_{t-1}(s_i),\>\>\>{\text{\normalfont with}}\>\omega, \alpha\ge0,
\end{align}
where $Z_{t}(s_j)$ is the $j$th component of a $N$-dimensional standardised generalised hyperbolic distribution, $\sigma^2_{t}(s_i,s_j)$ (fully defined in Section 3.1.2) is the covariance between $s_i$ and $s_j$, and $\sigma^2_{t}(s_i,s_i)$ is the spatially and temporally varying variance process. For simplicity, the autoregressive parameter is modelled as a single global effect $b$, whereas a more general structure can be introduced and is discussed in Section 5.\\
\indent Writing the above in vector and matrix notations, let $\boldsymbol Y_t=\{Y_t(s): s\in R\subset \mathbb{R}^N\}$ be a real-valued spatial process, we have the following spatio-temporal model, for $t=1,\dots,T$:
\begin{align}
\label{cen}
&\boldsymbol D_t=\boldsymbol A\boldsymbol Y_{t},\\
\label{meanp}
&\boldsymbol Y_{t}=\beta_0+\sum^M_{k=1}\boldsymbol X_{kt}\beta_k+\boldsymbol V_{t},\\
\label{sperror0}
&\boldsymbol V_t=b\boldsymbol V_{t-1}+\boldsymbol U_{t},\\
\label{sperror}
&\boldsymbol U_{t}=\boldsymbol\sigma_{t}\boldsymbol Z_{t},\\
\label{sigma22}
&\boldsymbol\sigma_t^2=\boldsymbol S_t\boldsymbol\rho\boldsymbol S_t,
\end{align}
where the matrices and vectors are defined in the following subsections.

%\eqref{cen} and \eqref{meanp} describe the latent mean process and \eqref{sperror0}, %\eqref{sperror} and \eqref{sigma22} characterise the spatially correlated errors. The  are defined %in the following subsections.
\subsubsection{The mean process}
 The mean process is modelled by \eqref{cen} and \eqref{meanp}. For a finite number of spatial locations $\left\{s_1,\dots,s_N\right\}\subset R$, the vector $\boldsymbol D_t=\left(D_t(s_1),\dots,D_t(s_N)\right)'$ is the observed precipitation at weather stations at time $t$, $\boldsymbol Y_t=\left(Y_t(s_1),\dots,Y_t(s_N)\right)'$ represents the latent underlying process at time $t$, $\boldsymbol X_{kt}=\left(X_{kt}(s_1),\dots,X_{kt}(s_N)\right)'$ for $k=1,\dots,M$ are the vector processes of $M$ known covariates at time $t$, and $\boldsymbol V_t=\left(V_t(s_1),\dots V_t(s_N)\right)'$ are the spatially correlated errors. We note that $\boldsymbol D_t, \boldsymbol Y_t, \boldsymbol X_{kt}, \boldsymbol V_t \in\mathbb{R}^{N}$ and that the $N\times N$ matrix $\boldsymbol A$ is a diagonal matrix whose diagonal entries starting in the upper left corner are given by standard indicator functions $ I_{\{ Y_{t}(s_i)\in E\}},\dots,I_{\{ Y_{t}(s_N)\in E\}}$. As we have already discussed, for simplicity $\beta_0\in\mathbb{R}$, $\beta_k\in\mathbb{R}$ are the global intercept and coefficient for the $k$th covariate respectively. \\
\indent We assume the spatial correlation matrix $\boldsymbol\rho$ has Mat{\'e}rn structure \citep{matern}, independent of time, i.e. the $(i,j)$ component of the matrix $\boldsymbol\rho$ is
\begin{align}
\label{cof}
\rho\left(s_i,s_j;\theta,\phi\right)=\frac{1}{2^{\phi-1}\Gamma{(\phi)}}\left(2\sqrt{\phi}\left\Vert s_i-s_j\right\Vert\theta\right)^\phi K_\phi\left(2\sqrt{\phi}\left\Vert s_i-s_j\right\Vert\theta\right),\>\>\>\theta,\phi>0,
\end{align}
where $\Gamma(\phi)$ is the standard gamma function, $K_\phi$ is the modified Bessel function of second kind with order $\theta$, and $\left\Vert s_i-s_j\right\Vert$ is the distance between $s_i$ and $s_j$. The parameter $\theta$ controls the rate of decay of the correlation as the distance $\left\Vert s_i-s_j\right\Vert$ increases and the parameter $\phi$ controls smoothness of the random field. Our covariance model assmes the process is isotropic and stationary; we found it impossible to detect any nonstationary or anisotropy with only 21 stations. For convenience we define one common variance $\sigma^2_\beta$ for all spatially varying parameters, i.e. $\sigma_k=\sigma_\beta$ for $k=1,\dots,M$.

\subsubsection{Spatially and temporally correlated errors}
The error process $\boldsymbol V_t$ is modelled by \eqref{sperror0}, \eqref{sperror} and \eqref{sigma22}, where $\boldsymbol Z_t\in\mathbb{R}^N$ is a standardised multivariate generalised hyperbolic random variable and $\boldsymbol\sigma_t$ is a lower trianglar $N\times N$ matrix given by the Cholesky decomposition of the $N\times N$ covariance process matrix $\boldsymbol\sigma^2_t$. In its covariance structure given in \eqref{sigma22}, the variance matrix $\boldsymbol S_t$ is a $N\times N$ diagonal matrix whose diagonal entries starting in the upper left corner are given by the ARCH process $\sqrt{\omega+\alpha\boldsymbol U^2_{t-1}}$ with $\omega,\alpha\in\mathbb{R}$, whereas the correlation matrix is given by $\boldsymbol\rho_t\in\mathbb{R}^{N\times N}$. In detail, we write
\begin{align}
\label{cp}
\boldsymbol\sigma^2_t=&\boldsymbol S_t\boldsymbol\rho\boldsymbol S_t\>\>\>{\text{\normalfont with}}\notag\\
%\begin{pmatrix}
%\sigma^2_t(s_1) & \sigma_t{(s_1)}\sigma_t{(s_2)}\rho_t{(s_1,s_2)} & \cdots & \sigma_t{(s_1)}\sigma_t{(s_N)}\rho_t{(s_1,s_N)} \\
%\sigma_t{(s_2)}\sigma_t{(s_1)}\rho_t{(s_2,s_1)} & \sigma^2_t(s_2) & \cdots &\sigma_t{(s_2)}\sigma_t{(s_N)}\rho_t{(s_2,s_N)}   \\
%\vdots & \vdots & \ddots & \vdots \\
%\sigma_t{(s_N)}\sigma_t{(s_1)}\rho_t{(s_N,s_1)} & \sigma_t{(s_N)}\sigma_t{(s_2)}\rho_t{(s_N,s_2)} & \cdots & \sigma^2_t(s_N)
%\end{pmatrix}\\
%\label{deco}
%&=
=&
\begin{pmatrix}
\sigma_t(s_1) & 0 & \cdots & 0 \\
0 & \sigma_t(s_2)  & \cdots & 0  \\
\vdots & \vdots & \ddots & \vdots \\
0 & 0& \cdots &\sigma_t(s_N) 
\end{pmatrix}
\begin{pmatrix}
1 & \rho(s_1,s_2) & \cdots & \rho{(s_1,s_N)} \\
\rho{(s_2,s_1)} & 1 & \cdots & \rho{(s_2,s_N)}  \\
\vdots & \vdots & \ddots & \vdots \\
\rho{(s_N,s_1)} & \rho{(s_N,s_2)} & \cdots & 1
\end{pmatrix}\notag\\
&
\begin{pmatrix}
\sigma_t(s_1) & 0 & \cdots & 0 \\
0 & \sigma_t(s_2)  & \cdots & 0  \\
\vdots & \vdots & \ddots & \vdots \\
0 & 0& \cdots &\sigma_t(s_N) 
\end{pmatrix},
\end{align}
where the spatial correlation matrix is given by \eqref{cof}. If we write
\begin{align}
{\text{\normalfont Chol}}
\begin{pmatrix}
1 & \rho(s_1,s_2) & \cdots & \rho{(s_1,s_N)} \\
\rho{(s_2,s_1)} & 1 & \cdots & \rho{(s_2,s_N)}  \\
\vdots & \vdots & \ddots & \vdots \\
\rho{(s_N,s_1)} & \rho{(s_N,s_2)} & \cdots & 1
\end{pmatrix}
=
\begin{pmatrix}
l_{11} & 0 & \cdots & 0 \\
l_{21} & l_{22} & \cdots & 0 \\
\vdots & \vdots & \ddots & \vdots \\
l_{N1} & l_{N2} & \cdots & l_{NN},
\end{pmatrix}
=\boldsymbol L
\end{align}
where $\boldsymbol L$ is a lower triangular matrix with
\begin{align}
l_{kk}=\sqrt{1-\sum^{k-1}_{j=1}l^2_{kj}}\>\>\>{\text{\normalfont and}}\>\>\>l_{ik}=\frac{1}{l_{kk}}\left(\rho(s_i,s_k)-\sum^{k-1}_{j=1}l_{ij}l_{kj}\right),
\end{align}
we have
\begin{align}
\boldsymbol\sigma_t=\boldsymbol S_t \boldsymbol L.
\end{align}

\subsubsection{Standardised generalised hyperbolic random variables}

The random vector $\boldsymbol Z_t\in\mathbb{R}^N$ follows a standardiased multivariate generalised hyperbolic distribution with parameters $\lambda, \psi,\chi\in\mathbb{R}$, $\boldsymbol\mu,\boldsymbol\tau\in\mathbb{R}^N$ and $\boldsymbol\Sigma\in\mathbb{R}^{N\times N}$. That is $\boldsymbol Z_{t}\sim MVGH\left(\lambda,\psi,\chi,\boldsymbol\mu, \boldsymbol\Sigma,\boldsymbol\tau\right)$ with
\begin{align}
\label{muvar}
\chi=1,\>\>\>\boldsymbol\mu=-L\left(\boldsymbol\tau,\psi,\lambda\right)\boldsymbol\tau\>\>\>{\text{\normalfont and}}\>\>\>\boldsymbol\Sigma=\frac{\sqrt{\psi}}{M_{\lambda+1}\left(\sqrt{\psi}\right)}\left(\boldsymbol I_N+\frac{L\left(\boldsymbol\tau,\psi,\lambda\right)-1}{\boldsymbol\tau'\boldsymbol\tau}\boldsymbol\tau\boldsymbol\tau'\right),
\end{align}
where $M_{\lambda+1}\left(\sqrt{\psi}\right)$ and $L\left(\boldsymbol\tau,\psi,\lambda\right)$ are given by \eqref{s1} and \eqref{s2}. Note that $\boldsymbol\tau$ is the $N\times1$ skewness parameter. In this paper we assume that $\tau(s_i)=m_\tau$ for $i=1,\dots,N$. We note that a spatial random effect on $\boldsymbol\tau$ can be introduced. However this might lead to a non-identifiability problem because there will be two sets of random effects on the overall skewness, including the covariance process $\boldsymbol\sigma^2_t$, only the product of which is identified by the data; see \eqref{a3}. Routine calculation then yields the conditional distribution for $\boldsymbol Y_t$:
\begin{align}
\label{disY}
\boldsymbol Y_t\vert\boldsymbol\zeta,\boldsymbol Y_{t-1}\sim MVGH\left(\lambda,\psi,1,\boldsymbol\mu_t, \boldsymbol\Sigma_t,\boldsymbol\tau_t\right), 
\end{align}
where $\boldsymbol\zeta=\left(a,b,\alpha,\omega,\lambda,\psi,\boldsymbol\tau,\beta_1,\dots,\beta_k,\theta\right)$ represents all parameters in the model and 
\begin{align}
\label{a1}
&\boldsymbol\mu_t=\beta_0+\sum^M_{k=1}\boldsymbol X_{kt}\boldsymbol\beta_k+b\boldsymbol V_{t-1}+\boldsymbol\sigma_t\boldsymbol\mu,\\
\label{a2}
&\boldsymbol\Sigma_t=\boldsymbol\sigma_{t}\boldsymbol\Sigma\boldsymbol\sigma'_t,\\
\label{a3}
&\boldsymbol\tau_t=\boldsymbol\sigma_{t}\boldsymbol\tau.
\end{align}

\subsection{Bayesian framework}
Inference for the parameters $\boldsymbol\zeta$ in our model given the underlying spatio-temporal process $\boldsymbol Y=(\boldsymbol Y_1,\dots,\boldsymbol Y_T)$ comes simply from Bayes rule:
\begin{align}
\label{bays}
p(\boldsymbol\zeta\vert \boldsymbol Y)\propto p(\boldsymbol Y\vert\boldsymbol\zeta)p(\boldsymbol\zeta),
\end{align}
where $p$ denotes a probability density.\\
\indent Since at any given time $t$, some components of $\boldsymbol Y_t$ may be censored, to compute the likelihood, we require a data argumentation procedure to recover censored $\boldsymbol Y_t$ components. To be precise, let $\boldsymbol Y_{O_t}$ where $O_t\subset\{1,\dots,N\}$ indicate the subsets of the process at time $t$ that occur above zero, whereas $\boldsymbol Y_{C_t}$ with $C_t=\{1,\dots,N\}\backslash O_t$ are the censored part of $\boldsymbol Y_t$, then the observed information $\mathscr{I}_t$ at time $t$ is
\begin{align}
\mathscr{I}_t=\{Y_t(s_i),i\in O_t\}\cup\{Y_t(s_i)<0,i\in C_t\}.
\end{align}
We note that $\boldsymbol D_t$ carries similar information regarding $\boldsymbol Y_t$ as $\mathscr{I}_t$, but we observe $0$ for $i \in C_t$ rather than only knowing $Y_t(s_i)<0$. The likelihood for  $\boldsymbol\zeta=\left(a,b,\alpha,\omega,\lambda,\psi,m_\tau,\beta_0,\boldsymbol\beta_1,\dots,\boldsymbol\beta_k,\theta\right)$ given $\mathscr{I}_t$ and $\boldsymbol Y_{t-1}$ is
\begin{align}
\label{da}
L(\boldsymbol\zeta\vert\mathscr{I}_t,\boldsymbol Y_{t-1})=p(\boldsymbol Y_{O_t}\vert\boldsymbol\zeta,\boldsymbol Y_{t-1})\int_{\boldsymbol y_t\le\boldsymbol0}p(\boldsymbol y_t\vert\boldsymbol Y_{O_t},\boldsymbol\zeta,\boldsymbol Y_{t-1})d\boldsymbol y_t,
\end{align}
which can be computed using a data augmentation method by embedding Monte Carlo integration within our Metropolis-Hastings algorithm. A similar method was also used in \cite{mypaper0} and \cite{de2005bayesian1}. 
\subsubsection{Prior distributions}
We assign priors to the model parameter $\boldsymbol\zeta$. We do not assume any prior knowledge on how the covariates are related to precipitation, and thus we choose uninformative priors for the parameters $\beta_1,\dots,\beta_M$ by considering $\beta_k\sim N(m,v)$, for all $k$, with zero mean and very large variance. In this application, we rely on knowledge of the space in which we model to set priors for the spatial decay parameter $\theta$ \citep{banerjee2004hierarchical,cooley2007bayesian}. Since we model in the latitude$\slash$longitude space, we use Unif$(0,1.5)$ as our prior, which sets the maximum range of the exponential variogram model to be approximately 200 km.\\
\indent We also assign uninformative priors to the parameters of the generalised hyperbolic processes similar to the priors for $\beta_k$. A sensitivity analysis has also been done using different hyper-parameter values of the prior distributions; see details in Section 4.2.
\subsubsection{Sampling from the posterior}
For each $t\in\{1,\dots,T\}$, the hierarchical model described above yields the following posterior density 
\begin{align}
p(\boldsymbol Y_{C_t},\boldsymbol\zeta\vert\mathscr{I}_t,\boldsymbol Y_{t-1})\propto& p\left(\mathscr{I}_t,\boldsymbol Y_{C_t},\boldsymbol\zeta\vert\boldsymbol Y_{t-1}\right)\propto p(\mathscr{I}_t\vert\boldsymbol Y_{C_t},\boldsymbol\zeta,\boldsymbol Y_{t-1}) p(\boldsymbol Y_{C_t}\vert\boldsymbol\zeta)p(\boldsymbol\zeta)\notag\\
\propto&\frac{\left(\sqrt{\psi/\chi}\right)^\lambda\left(\psi+{\boldsymbol\gamma}'_t{\boldsymbol\Sigma}_t^{-1}{\boldsymbol\gamma}_t\right)^{\frac{N}{2}-\lambda}}{\left(2\pi\right)^{\frac{N}{2}}\left\vert{\boldsymbol\Sigma}_t\right\vert^{\frac{1}{2}}K_\lambda\left(\sqrt{\psi/\chi}\right)}\notag\\
&\times\frac{K_{\lambda-\frac{N}{2}}\left(\sqrt{\left(\chi+Q\left(\boldsymbol Y_t\right)\right)\left(\psi+{\boldsymbol\gamma}'_t{\boldsymbol\Sigma}_t^{-1}{\boldsymbol\gamma_t}\right)}\right)\exp\left(\boldsymbol Y_t-{\boldsymbol\mu}_t\right)'{\boldsymbol\Sigma}_t^{-1}{\boldsymbol\gamma}_t}{\left(\sqrt{\left(\chi+Q\left(\boldsymbol Y_t\right)\right)\left(\psi+{\boldsymbol\gamma}'_t{\boldsymbol\Sigma}_t^{-1}{\boldsymbol\gamma}_t\right)}\right)^{\frac{N}{2}-\lambda}}\notag\\
&\times\prod_{i\in C_t}\boldsymbol1_{\{Y_t(s_i)<0\}}\times\pi\left(b\right)\pi\left(\beta_0\right)\prod^M_{k=1}\pi\left(\beta_k\right)\pi\left(\omega\right)\pi\left(\alpha\right)\pi\left(\lambda,\psi\right)\pi\left(m_\tau\right)\pi\left(\theta\right)\pi\left(\phi\right),
\end{align}
where $Q\left(\boldsymbol Y_t\right)=\left(\boldsymbol Y_t-{\boldsymbol\mu}_t\right)'{\boldsymbol\Sigma}_t^{-1}\left(\boldsymbol Y_t-{\boldsymbol\mu}_t\right)$, $\boldsymbol\gamma_t=\boldsymbol\Sigma_t\boldsymbol\tau_t$, and $\boldsymbol\mu_t$, $\boldsymbol\Sigma_t$ and $\boldsymbol\tau_t$ are given by \eqref{a1}, \eqref{a2} and \eqref{a3}. We note that $\boldsymbol Y_t=(\boldsymbol Y_{O_t},\boldsymbol Y_{C_t})$ is the underlying vector, whose censored components $\{Y_t(s_i),i\in C_t\}$ are initialised at the censoring limit zero and then sampled collectively at each iteration of the Metropolis-Hastings algorithm from their full conditional distribution. Here we assume that $\boldsymbol Y_{C_{t-1}}$ has been recovered, so $\boldsymbol Y_{t-1}$ is avaliable. \\
\indent  Let $\boldsymbol\zeta^{[j]}=\left(a^{[j]},b^{[j]},\alpha^{[j]},\omega^{[j]},\lambda^{[j]},\psi^{[j]},m_\tau^{[j]},\beta^{[j]}_0, \beta^{[j]}_1,\dots,\beta^{[j]}_k,\theta^{[j]},\phi^{[j]}\right)$ be the $j$th realisation in the parameter space. We can rewrite \eqref{disY} as $\boldsymbol Y_t\vert\boldsymbol\zeta^{[j]}=\left(\boldsymbol Y_{O_t},\boldsymbol Y_{C_t}\right)\vert\boldsymbol\zeta^{[j]}\sim MGH\left(\lambda^{[j]},\psi^{[j]},1,\boldsymbol\mu_t^{[j]}, \boldsymbol\Sigma_t^{[j]},\boldsymbol\tau_t^{[j]}\right)$, where $\boldsymbol\mu_t^{[j]}$, $\boldsymbol\tau_t^{[j]}$  and $\boldsymbol\Sigma_t^{[j]}$ are given by \eqref{a1}, \eqref{a2} and \eqref{a3} with parameters $\boldsymbol\zeta^{[j]}$. If we write 
\begin{align}
\boldsymbol\mu_t^{[j]}=
\begin{pmatrix}
{\boldsymbol\mu}_{O_t}  \\
{\boldsymbol\mu}_{C_t} 
\end{pmatrix},\>
\boldsymbol\tau_t^{[j]}=
\begin{pmatrix}
{\boldsymbol\tau}_{O_t}  \\
{\boldsymbol\tau}_{C_t} 
\end{pmatrix}\>{\text{\normalfont and}}\>\>
\boldsymbol\Sigma_t^{[j]}=
\begin{pmatrix}
{\boldsymbol\Sigma}_{O_t} & {\boldsymbol\Sigma}_{{OC}_t}\\
{\boldsymbol\Sigma}_{{CO}t} & {\boldsymbol\Sigma}_{C_t}
\end{pmatrix},
\end{align}
then the censored components of $\boldsymbol Y_t$ given $\boldsymbol\zeta^{[j]}$ and the observed information $\mathscr{I}_t$ follow a multivariate GH distribution:
\begin{align}
\label{zero}
p(\boldsymbol Y_{C_t}\vert\boldsymbol\zeta^{[j]},\mathscr{I}_t)=&p(\boldsymbol Y_{C_t}\vert\boldsymbol\zeta^{[j]},\boldsymbol Y_{O_t})\notag\\
=&MVGH\left(\lambda',\chi',\psi',\boldsymbol\mu'_t,\boldsymbol\Sigma'_t,\boldsymbol\gamma'_t\right)\times\boldsymbol1_{\{\boldsymbol Y_{t}<\boldsymbol0\}}.
\end{align}
Here
\begin{align}
&\lambda'=\lambda-\dim{\left(\boldsymbol Y_{O_t}\right)}/2\\
&\boldsymbol\mu'_t=\boldsymbol\mu_{C_t}+\left(\boldsymbol Y_{O_t}-\boldsymbol\mu_{O_t}\right)\boldsymbol\Sigma_{O_t}^{-1}\boldsymbol\Sigma_{{OC}_t}\\
&\boldsymbol\Sigma'_t=\left\vert\boldsymbol\Sigma_{O_t}\right\vert^{1/\dim(\boldsymbol Y_{C_t})}\left(\boldsymbol\Sigma_{C_t}-\boldsymbol\Sigma_{{CO}_t}\boldsymbol\Sigma_{C_t}^{-1}\boldsymbol\Sigma_{{OC}_t}\right)\\
&\boldsymbol\gamma'_t=\left(\boldsymbol\gamma_{C_t}-\boldsymbol\gamma_{O_t}\boldsymbol\Sigma_{O_t}^{-1}\right)\left(\boldsymbol\Sigma_{C_t}-\boldsymbol\Sigma_{{CO}_t}\boldsymbol\Sigma_{O_t}^{-1}\right)^{-1}\boldsymbol\Sigma'_t\\
&\chi'=\left\vert\boldsymbol\Sigma_{O_t}\right\vert^{1/\dim(\boldsymbol Y_{C_t})}\left(1+\left(\boldsymbol Y_{O_t}-\boldsymbol\mu_{O_t}\right)\boldsymbol\Sigma_{O_t}^{-1}\left(\boldsymbol Y_{O_t}-\boldsymbol\mu_{O_t}\right)'\right)\\
&\psi'=\left\vert\boldsymbol\Sigma_{O_t}\right\vert^{1/N}\left(\psi+\boldsymbol\gamma'_{O_t}\boldsymbol\Sigma_{O_t}^{-1}\boldsymbol\gamma_{O_t}\right)\left\vert\boldsymbol\Sigma_{O_t}\right\vert^{1/\dim(\boldsymbol Y_{C_t})}-\\
&\left(\left(\boldsymbol\gamma_{C_t}-\boldsymbol\gamma_{O_t}\boldsymbol\Sigma_{O_t}^{-1}\right)\left(\boldsymbol\Sigma_{C_t}-\boldsymbol\Sigma_{{CO}_t}\boldsymbol\Sigma_{O_t}^{-1}\right)^{-1}\right)'\boldsymbol\Sigma'_t\left(\boldsymbol\gamma_{C_t}-\boldsymbol\gamma_{O_t}\boldsymbol\Sigma_{O_t}^{-1}\right)\left(\boldsymbol\Sigma_{C_t}-\boldsymbol\Sigma_{{CO}_t}\boldsymbol\Sigma_{O_t}^{-1}\right)^{-1},
\end{align}
with $\boldsymbol\gamma_{O_t}=\boldsymbol\Sigma_{O_t}\boldsymbol\tau_{O_t}+\boldsymbol\Sigma_{{OC}_t}\boldsymbol\tau_{C_t}$ and $\boldsymbol\gamma_{C_t}=\boldsymbol\Sigma_{{CO}_t}\boldsymbol\tau_{O_t}+\boldsymbol\Sigma_{{C}_t}\boldsymbol\tau_{C_t}$. See Section 3.3 for further details.

Data augmentation is done for all $t$ in a recursive manner, so $\boldsymbol Y_{C_t}$ is recovered and $\boldsymbol Y_t$ is available for all $t$. This method for censored observations ensures that likelihood distributions given data $\mathscr{I}_t$ follow \eqref{da} \citep{de2005bayesian1}. Under the assumption that $\boldsymbol Y_{t}$ and $\left(\boldsymbol Y_{C_t},\mathscr{I}_t\right)$ carry the same information, the posterior density has the form
\begin{align}
p\left(\boldsymbol Y_{C_T},\boldsymbol Y_{C_{T-1}},\dots,\boldsymbol Y_{C_1},\boldsymbol\zeta\vert\mathscr{I}_T,\mathscr{I}_{T-1},\dots,\mathscr{I}_1\right)=&\prod_{t=1}^Tp\left(\boldsymbol Y_{C_t},\boldsymbol\zeta\vert\mathscr{I}_t,\boldsymbol Y_{C_{t-1}},\mathscr{I}_{t-1}\right)\times p\left(\boldsymbol Y_{C_0},\mathscr{I}_0\right)\notag\\
=&\prod_{t=1}^Tp\left(\boldsymbol Y_{C_t},\boldsymbol\zeta\vert\mathscr{I}_t,\boldsymbol Y_{t-1}\right)\times p\left(\boldsymbol Y_{0}\right).
\end{align}
Due to the recursive nature of the hierarchical model, full conditionals are non-standard and require Metropolis-Hastings sampling. See details in \cite{gelman2013bayesian} and \cite{tierney1994markov}.

\subsection{Spatial predictive distribution}
Recall that our primary goal is to make spatial predictions for out-of-sample locations given the observed data. In practice posterior predictive distributions can be produced by sampling from a finite set of locations $\boldsymbol s^*$, conditional on $\boldsymbol Y$ and $\boldsymbol\zeta$. We want to predict $\boldsymbol Y^*_{t}=\left(Y_t(s^*_{1}),\dots,Y_t(s^*_{r})\right)$ at locations $\boldsymbol s^*=\left\{s^*_{1},\dots,s^*_{r}\right\}$ at time $t$. Let us consider the following hierarchical model:
\begin{align}
&\widetilde{\boldsymbol Y}_{t}=\beta_0+\sum^M_{k=1}\widetilde{\boldsymbol X}_{kt}{\beta}_k+\widetilde{\boldsymbol V}_{t},\\
&\widetilde{\boldsymbol V}_t=b\widetilde{\boldsymbol V}_{t-1}+\widetilde{\boldsymbol U}_{t},\\
&\widetilde{\boldsymbol U}_{t}=\widetilde{\boldsymbol\sigma}_{t}\widetilde{\boldsymbol Z}_{t},\\
\label{sigma2}
&\widetilde{\boldsymbol\sigma}_t^2=\widetilde{\boldsymbol S}\widetilde{\boldsymbol\rho}_t\widetilde{\boldsymbol S},
\end{align}
where 
\begin{align}
\widetilde{\boldsymbol Y}_t=
\begin{pmatrix}
\boldsymbol Y_t  \\
\boldsymbol Y_t^*
\end{pmatrix},
\widetilde{\boldsymbol X}_{kt}=
\begin{pmatrix}
\boldsymbol X_{kt}  \\
\boldsymbol X_{kt}^*
\end{pmatrix},
%\widetilde{\boldsymbol\beta}_k=
%\begin{pmatrix}
%\boldsymbol \beta_k  \\
%\boldsymbol\beta_k^*
%\end{pmatrix},
\widetilde{\boldsymbol V}_t=
\begin{pmatrix}
\boldsymbol V_t  \\
\boldsymbol V_t^*
\end{pmatrix},
\widetilde{\boldsymbol U}_t=
\begin{pmatrix}
\boldsymbol U_t  \\
\boldsymbol U_t^*
\end{pmatrix},\>{\text{\normalfont and}}\>
\widetilde{\boldsymbol Z}_t=
\begin{pmatrix}
\boldsymbol Z_t  \\
\boldsymbol Z_t^*
\end{pmatrix},
\end{align}
where $\widetilde{\boldsymbol Z}_t\sim MGH\left(\lambda,\psi,\chi,\widetilde{\boldsymbol\mu}, \widetilde{\boldsymbol\Sigma},\widetilde{\boldsymbol\tau}\right)$ satisfying the standardisation condition. The generalised hyperbolic parameters $\lambda, \chi$ and $\psi$ are the same as before, and $\widetilde{\boldsymbol\mu}, \widetilde{\boldsymbol\Sigma}$ and $\widetilde{\boldsymbol\tau}$ are given by \eqref{muvar}. We have $\widetilde{\boldsymbol Y}_t, \widetilde{\boldsymbol X}_t, \widetilde{\boldsymbol V}_t, \widetilde{\boldsymbol U}_t, \widetilde{\boldsymbol Z}_t\in\mathbb{R}^{N+r}$ and $b,\beta_0,\beta_k\in\mathbb{R}$. The conditional variance $\widetilde{\boldsymbol\sigma}^2_t$ of $\widetilde{\boldsymbol U}_t$ becomes
\begin{align}
\widetilde{\boldsymbol\sigma}^2_t=
\begin{pmatrix}
\boldsymbol\sigma_t^2 & \boldsymbol\sigma_t{\left(\boldsymbol s^*,\boldsymbol s\right)} \\
\boldsymbol\sigma_t{\left(\boldsymbol s,\boldsymbol s^*\right)}  &    \boldsymbol\sigma_t^{*2}
\end{pmatrix},
\end{align}
and consequently,
\begin{align}
\widetilde{\boldsymbol\sigma}_t=\widetilde{\boldsymbol S}\>{\text{\normalfont Chol}}\left(\widetilde{\boldsymbol\rho}\right)=
\begin{pmatrix}
\boldsymbol\sigma_t\left(\boldsymbol s\right) & \boldsymbol 0 \\
\boldsymbol 0  &    \boldsymbol\sigma_t\left(\boldsymbol s^*\right) 
\end{pmatrix}
{\text{\normalfont Chol}}
\begin{pmatrix}
\boldsymbol\rho & \boldsymbol\rho{\left(\boldsymbol s^*,\boldsymbol s\right)} \\
\boldsymbol\rho{\left(\boldsymbol s,\boldsymbol s^*\right)}  &    \boldsymbol\rho^{*},
\end{pmatrix}
\end{align}
where $\boldsymbol\sigma_t\left(\boldsymbol s^*\right)\in\mathbb{R}^{r\times r}$ is a diagonal matrix whose entries are given by $\sqrt{\omega+\alpha\boldsymbol U^{*2}_{t-1}}$, and $\boldsymbol\rho{\left(\boldsymbol s^*,\boldsymbol s\right)}\in\mathbb{R}^{N\times r}$, $\boldsymbol\rho{\left(\boldsymbol s,\boldsymbol s^*\right)}\in\mathbb{R}^{r\times N}$ and $\boldsymbol\rho^{*}\in\mathbb{R}^{r\times r}$ are given by \eqref{cof}. Hence we have
\begin{align}
\label{sppred}
\widetilde{\boldsymbol Y}_t\vert \boldsymbol\zeta^\dagger,\widetilde{\boldsymbol Y}_{t-1},=
\begin{pmatrix}
\boldsymbol Y_t  \\
\boldsymbol Y_t^*
\end{pmatrix}\vert
\begin{pmatrix}
\boldsymbol Y_{t-1}  \\
\boldsymbol Y_{t-1}^*
\end{pmatrix}\sim MGH\left(\lambda,\chi,\psi,\boldsymbol\mu^\dagger_t,\boldsymbol\Sigma^\dagger_t,\boldsymbol\gamma^\dagger_t\right),
\end{align}
where $\boldsymbol\zeta^\dagger$ represents all parameters in the model, $\boldsymbol\gamma^\dagger_t=\boldsymbol\Sigma^\dagger_t\boldsymbol\tau^\dagger_t=(\boldsymbol\gamma_{1t}\>\>\boldsymbol\gamma_{0t})'$, and 
\begin{align}
\boldsymbol\mu^\dagger_t=\beta_0+\sum^M_{k=1}\widetilde{\boldsymbol X}_{kt}\beta_k+b\widetilde{\boldsymbol V}_{t-1}+\widetilde{\boldsymbol\sigma}_t\widetilde{\boldsymbol\mu}=
\begin{pmatrix}
{\boldsymbol\mu}_{1t}  \\
{\boldsymbol\mu}_{0t} 
\end{pmatrix},\>
\boldsymbol\tau^\dagger_t=\widetilde{\boldsymbol\sigma}_t\widetilde{\boldsymbol\tau}=
\begin{pmatrix}
{\boldsymbol\tau}_{1t}  \\
{\boldsymbol\tau}_{0t} 
\end{pmatrix}\>{\text{\normalfont and}}\>\>\boldsymbol\Sigma^\dagger_t=\widetilde{\boldsymbol\sigma}_t\widetilde{\boldsymbol\Sigma}\widetilde{\boldsymbol\sigma}'_t
=\begin{pmatrix}
{\boldsymbol\Sigma}_{1t} & {\boldsymbol\Sigma}_{10t}\\
{\boldsymbol\Sigma}_{01t} & {\boldsymbol\Sigma}_{0t},
\end{pmatrix}.\notag
\end{align}
Due to the recursive temporal structure of the hierarchical model, we assume $\boldsymbol Y_{t-1}^*$ is sampled in previous step and is known. We also assume that censored components of $\boldsymbol Y_t$ and $\boldsymbol Y_{t-1}$ are recovered using \eqref{zero}. Here we have ${\boldsymbol\mu}_{1t}, {\boldsymbol\gamma}_{1t}\in\mathbb{R}^{N}$, ${\boldsymbol\mu}_{0t}, {\boldsymbol\gamma}_{0t}\in\mathbb{R}^{r}$, ${\boldsymbol\Sigma}_{1t}\in\mathbb{R}^{N\times N}$, ${\boldsymbol\Sigma}_{0t}\in\mathbb{R}^{r\times r}$, ${\boldsymbol\Sigma}_{10t}\in\mathbb{R}^{N\times r}$ and ${\boldsymbol\Sigma}_{01t}\in\mathbb{R}^{r\times N}$. The subscript $1$ indicates observed locations and $0$ is the location(s) we want  to predict. Finally we use the algorithm outlined in Section 3.2 to obtain MCMC samples for the posteriors of the parameters in \eqref{sppred}. To obtain the spatial predictive distribution, using the result in \cite{hardle2007applied}, we have 
\begin{align}
\label{sppp}
\boldsymbol Y_t^*\vert \boldsymbol Y_t=\boldsymbol y_{t}\sim MGH\left(\lambda',\chi',\psi',\boldsymbol\mu'_t,\boldsymbol\Sigma'_t,\boldsymbol\gamma'_t\right),
\end{align}
where
\begin{align}
&\lambda'=\lambda-\dim{\left(\boldsymbol Y_{1t}\right)}/2\\
&\boldsymbol\mu'_t=\boldsymbol\mu_{0t}+\left(\boldsymbol y_{1t}-\boldsymbol\mu_{1t}\right)\boldsymbol\Sigma_{1t}^{-1}\boldsymbol\Sigma_{10t}\\
&\boldsymbol\Sigma'_t=\left\vert\boldsymbol\Sigma_{1t}\right\vert^{1/\dim(\boldsymbol Y_{0t})}\left(\boldsymbol\Sigma_{0t}-\boldsymbol\Sigma_{01t}\boldsymbol\Sigma_{0t}^{-1}\boldsymbol\Sigma_{10t}\right)\\
&\boldsymbol\gamma'_t=\left(\boldsymbol\gamma_{0t}-\boldsymbol\gamma_{1t}\boldsymbol\Sigma_{1t}^{-1}\right)\left(\boldsymbol\Sigma_{0t}-\boldsymbol\Sigma_{01t}\boldsymbol\Sigma_{1t}^{-1}\right)^{-1}\boldsymbol\Sigma'_t\\
&\chi'=\left\vert\boldsymbol\Sigma_{1t}\right\vert^{1/\dim(\boldsymbol Y_{0t})}\left(\chi+\left(\boldsymbol y_{1t}-\boldsymbol\mu_{1t}\right)\boldsymbol\Sigma_{1t}^{-1}\left(\boldsymbol y_{1t}-\boldsymbol\mu_{1t}\right)'\right)\\
&\psi'=\left\vert\boldsymbol\Sigma_{0t}\right\vert^{1/N}\left(\psi+{\boldsymbol\gamma}'_{0t}\boldsymbol\Sigma^{-1}_{0t}\boldsymbol\gamma_{0t}\right)\left\vert\boldsymbol\Sigma_{1t}\right\vert^{1/\dim(\boldsymbol Y_{0t})}-\\
&\left(\left(\boldsymbol\gamma_{0t}-\boldsymbol\gamma_{1t}\boldsymbol\Sigma_{1t}^{-1}\right)\left(\boldsymbol\Sigma_{0t}-\boldsymbol\Sigma_{01t}\boldsymbol\Sigma_{1t}^{-1}\right)^{-1}\right)'\boldsymbol\Sigma'_t\left(\boldsymbol\gamma_{0t}-\boldsymbol\gamma_{1t}\boldsymbol\Sigma_{1t}^{-1}\right)\left(\boldsymbol\Sigma_{0t}-\boldsymbol\Sigma_{01t}\boldsymbol\Sigma_{1t}^{-1}\right)^{-1}.
\end{align}
We also used this result for \eqref{zero}.

\subsection{Temporal predictive distribution}
In this subsection we briefly describe one-step ahead temporal prediction. To obtain one-step ahead temporal predictions, or forecasts, we consider two particular cases; forecasting in the observed spatial location $s_i$, $i=1,\dots,N$, and in out-of-sample spatial location $ s_0$.\\
\indent Suppose we want to forecast at time $T+1$ at location $s_i$ denoted by $D_{T+1}(s_i)$. We can easily obtain the posterior temporal predictive distribution of $D_{T+1}(s_i)$ from that of $Y_{T+1}(s_i)$, which can be calculated via
\begin{align}
Y_{T+1}(s_i)=\beta_0+\sum^M_{k=1} \beta_kX_{k(T+1)}(s_i)+ V_{T+1}(s_i),
\end{align}
where
\begin{align}
&V_{T+1}(s_i)=bV_{T}(s_{i})+U_{T+1}(s_i)\>\>\>{\text{\normalfont and}}\notag\\
&U_{T+1}(s_i)=\sum^N_{j=1}\sigma_{T+1}(s_i,s_j)Z_{T+1}(s_j)\>\>\>{\text{\normalfont with}}\>\>\>\sigma^2_{T+1}\left(s_i,s_i\right)=\omega+\alpha U^2_{T}(s_i).
\end{align}
We note that the correlation structure of the error term $U$ and the distributional property of $Z$ are independent of time.\\
\indent For forecasting in a new location $s_0$ at time $T+1$ we define the observation $D_{T+1}(s_0)$. We already obtain the forecast values $D_{T+1}(s_i)$ and hence to obtain the forecast distribution in location $s_0$, we use the spatial predictive distribution described in the previous section.

\section{Modelling results}
The WA winter daily rainfall data described in Section 2 are analysed using our Bayesian hierarchical model. For comparision, two models are fitted: one for winter daily precipitation in years 1988-2008, and the other for winter weekly precipitation totals in year 1958-2008. Both models are fitted with 10,000 MCMC iterations, and the first 3,000 samples are discarded as burn-in. The MCMC chains converged quickly within a few hundred iterations. For brevity, these results and other MCMC diagnostics (see, \cite{gelman1992inference}) are omitted. 
\subsection{Model-based analysis}
Table 2 shows estimates of the parameters of the daily precipitation model. The autoregressive parameter $b$ for the errors show a significant positive autocorrelation between precipitation on successive days. A similar positive autocorrelation is shown for the variance model. Summary statistics for the three GH parameters are also shown below. A small positive value for $m_\tau$ reflects the fact that the data is substaintially right-skewed. For simplicity, we have used the exponetional covariance function in our model so the smoothing parameter $\phi=0.5$. The median of the spatial decay parameter $\theta$ is estimated to be 1.03, and the 95\% CI varies from 0.97 to 1.09. This corresponds to an approximate 291 km spatial effective range of dependency over the study region. Summary statistics for the global effect of the four covariates are also shown in Table 2. As expected, elevation shows a negative effect on precipitation for our study region. We also find that the global effect of the covariate NINO 3.4 is positive and statistically significant, consistent with results in \cite{crimp2014bayesian}.\\
\begin{table}[h] 
\footnotesize
\centering
\begin{tabular}{lllllll}
\toprule 
\toprule 
Parameters &  Median & Std.dev  & 2.5\% & 97.5\% \\ 
    \midrule
$b$ (AR) & 5.31$\times 10^{-2}$& 2.44$\times 10^{-2}$& 9.52$\times 10^{-3}$&1.05$\times 10^{-1}$\\ 
$\omega$ (ARCH) & 131.81 & 14.56 & 110.88 &165.80\\ 
$\alpha$ (ARCH) & 1.44$\times 10^{-2}$ & 6.27$\times 10^{-3}$  &3.84$\times 10^{-3}$ &2.90$\times 10^{-2}$\\ 
$\psi$ (Shape) & 5.99$\times 10^{-3}$ & 1.15$\times 10^{-3}$  & 4.15$\times 10^{-3}$&8.62$\times 10^{-3}$  \\ 
$\lambda$ (Subclass)&6.25$\times 10^{-2}$&6.26$\times 10^{-2}$   & -6.61$\times 10^{-2}$ &1.76$\times 10^{-1}$\\ 
$m_\tau $ (Skewness) & 3.67$\times 10^{-4}$& 5.03$\times 10^{-4}$  & 1.49$\times 10^{-5}$&1.87$\times 10^{-3}$\\ 
$\theta$ (Spatial decay) &1.03 & 3.03$\times 10^{-2}$ & 9.73$\times 10^{-1}$&1.09\\ 
\\
$\beta_0$ (Global intercept) & 1.09 & 7.79$\times 10^{-2}$ &9.45$\times 10^{-1}$ & 1.25 \\ 
$\beta_1$ (NINO 3.4) & 7.61$\times 10^{-2}$ &5.73$\times 10^{-2}$  & 3.44$\times 10^{-3}$ &1.86$\times 10^{-1}$ \\ 
$\beta_2$ (SAMI) & -1.19$\times 10^{-1}$ &3.66$\times 10^{-2}$  & -1.57$\times 10^{-1}$ &-7.89$\times 10^{-2}$ \\ 
$\beta_3$ (IOD) & 3.11$\times 10^{-1}$ &5.28$\times 10^{-1}$  & -2.53$\times 10^{-1}$ &6.94$\times 10^{-1}$ \\ 
$\beta_4$ (ELE) &-3.14$\times 10^{-2}$  &4.63$\times 10^{-3}$ &-1.21$\times 10^{-1}$&1.06$\times 10^{-2}$ \\
    \bottomrule
\end{tabular}
 \label{restab} % is used to refer this table in the text 
\caption{\footnotesize Posterior mean and corresponding 95\% CIs of the parameters of the spatio-temporal model for the study region.} % title of Table 
\end{table} 
\indent For brevity, results for the weekly precipitation model is not presented here. The effect of the four covariates on weekly precipitation totals is similar to those in Table 2. However, other parameter estimates vary considerably reflecting different skewness, temporal and spatial dependencies in the weekly data. For instance, the median of the spatial decay parameter $\theta$ is estimated to be $0.43$, and the 95\% CI varies from $0.28$ to $0.56$, corresponding to a much larger spatial effective range of dependency (697 km) in weekly precipitation totals over the study region.

\subsection{Sensitivity analysis}
One of the disadvantages of Bayesian analysis is due to the fact that the prior distribution for the parameters can have a significant impact on the posterior distribution, and consequently, lead to biased results. We checked the sensitivity of the model by using diffierent  hyper-parameters of the Normal and truncated Normal priors. In the case where the original priors are uniform distributions, Beta distributions on the same support are used as alternatives. The results showed that our model is not very sensitive to the choice of the hyper-parameter values. For brevity these results have been omitted from the paper.
\subsection{Spatial predictions of daily and weekly precipitation and its extremes}
To check and validate the model's performances, we produce spatial predictions of precipitation return periods at out-of-sample locations, and compare them to empirical return preiods. As noted previously, the calculation of return period assumes that the probability of the event occurring does not vary over time and is independent of past event. Although this is only true for large precipitation, we still calculate it for all observed precipitation amounts solely for model validation purposes.\\
\indent To produce out-of-sample spatial predictions for return periods, we hold back one weather station at a time and use the previously outlined algorithm and \eqref{sppp} to obtain samples from the spatial predictive distribution for that location, then calculate return periods using the method described in Section 1.1. Finally the 95\% prediction credible interval (CI) for the return period curve is produced based on 1,000 simulations. The left panel of Figure 2 shows the spatial prediction of daily precipitation return periods at Site 9619. The red dots are empirical return periods, whereas the dashed lines represent the 95\% CI. We can also compare samples from the spatial predictive distribution directly to observed daily precipitation through a Q-Q plot as shown in the right panel of Figure 2. Observed quantiles of winter daily precipitation observed in year 1988-2008 are on the x-axis, and mean simulated quantiles are on the y-axis. The same two graphs for winter weekly precipitation totals in years 1958-2008 are also produced at Site 9619. The model appears to produce excellent spatial predictions, including for extreme precipitation events, and accommodate different shapes and skewness. It is important to recognise that the CIs for return periods constructed under the proposed model are much narrower than those constructed under extreme value theory, e.g. \cite{li2005statistical}. Out-of-sample predictions of return periods and the Q-Q plots for both daily and weekly precipitation data at all other weather stations are presented in Figure 7, 8, 9 and 10 in the Appendix. Similar conclusions can be drawn from those plots. Given the demanding criteria of accurate out-of-sample prediction of extreme precipitation, these results are quite impressive. However, in Figure 2, the out-of-sample spatial prediction for winter daily precipitation return periods shows a noticible misalignment at 0 mm and for very small precipitation amounts. This might be due to model error or measurement errors of small precipitation amounts.\\
\begin{figure}[h]
 \centerline{\includegraphics[width=195mm]{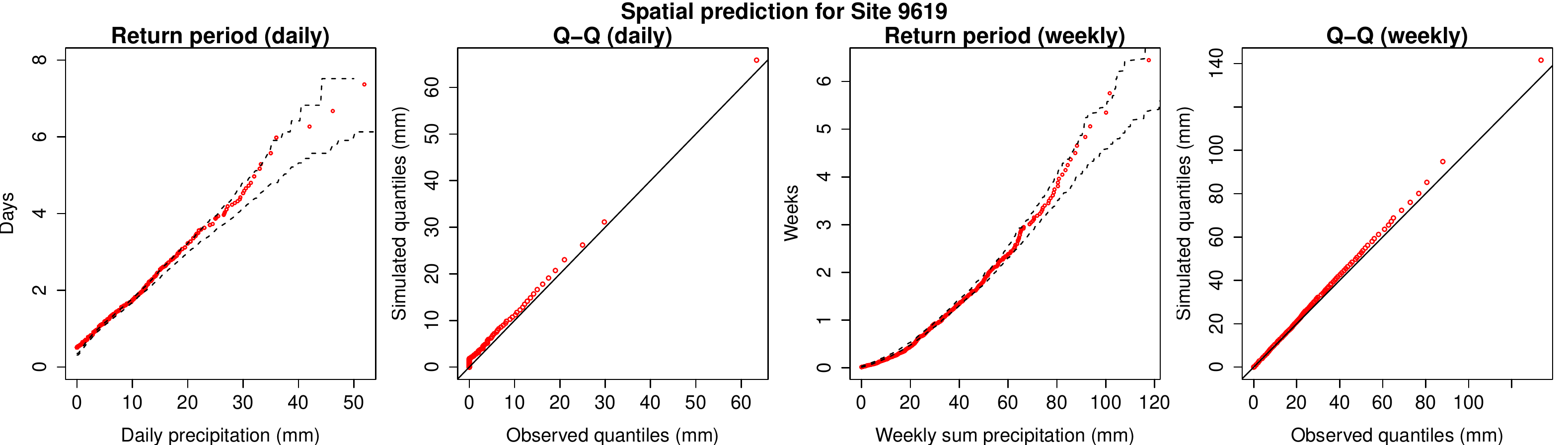}}
 \caption{\footnotesize Out-of-sample spatial prediction of daily and weekly sum precipitation for Site 9619, based on 1,000 simulations.}
\end{figure}
\indent Figure 3 shows the empirical and modelled spatio-temporal variograms for both daily and weekly precipitation data. For both daily and weekly data, the modelled variogram closely matches the empirical variogram. Also it is worth noting that total weekly precipitation shows much less temporal dependency than daily precipitation. Furthermore, there is an increase in the variogram for lag 1 and greater for distance close to zero, which is expected for rainfall data.\\
\begin{figure}[h]
 \centerline{\includegraphics[width=150mm]{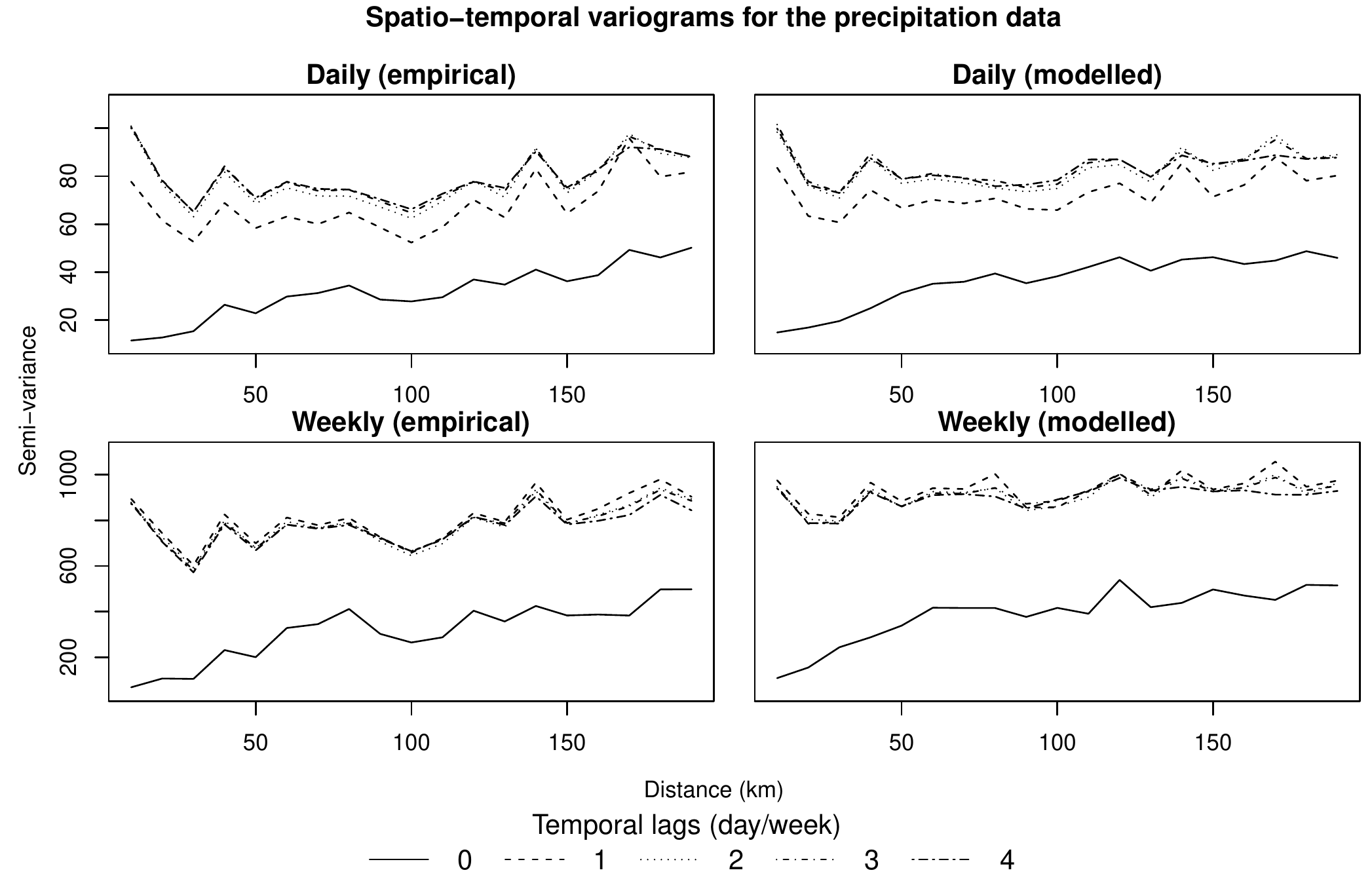}}
 \caption{\footnotesize Empirical and modelled spatio-temporal variograms for both daily and weekly precipitation data. Modelled variograms are based on 1,000 simulations.}
\end{figure}
\indent Another frequently used uncertainty measure for precipitation and its extremes is precipitation duration. Figure 4 shows out-of-sample spatial predictions of mean duration (days) over/below thresholds for winter daily precipitation. The box-plots are based on 1,000 simulations, and the red dots represent observed mean durations (days) over/below thresholds for 1988-2008. It shows that the model matches observations at most out-of-sample locations, especially for duration over large thresholds, indicating that the model efficiently represents spatial and short-term temporal dependencies. Similarly, out-of-sample spatial predictions of numbers of weeks over/below thresholds are presented in Figure 5, with the top panel showing spatial predictions of the number of weeks with no rainfall at each out-of-sample weather station. The red dots represent observed numbers of winter weeks over/below thresholds for 1958-2008. Again the out-of-sample predictions are in close agreement with the observed data.\\
\indent Finally, we produce predictive maps for numbers of weeks over/below thresholds. For prediction purposes, we define 150 grid points over our study region and obtained interpolated predicted values of the winter weekly precipitation totals using data from the 21 weather stations for the time period 1958-2008. Figure 6 shows the predicted mean spatial pattern of the number of weeks with aggregated precipitation greater than 90 mm and exactly 0 mm, respectively. The values that appear on the map represent the observed values. The plots are accompanied by their standard deviation plots. The model shows close agreement with observations across most of the study region. 

\section{Discussion}
In this paper, we developed a Bayesian heirarchical model that utilises the generalised hyperbolic process for producing spatial predictions and measures of uncertainty for spatio-temporal data that is heavy-tailed and subject to substantial and varying skewness. Unlike models based on extreme value theory, which only model maxima of finite-sized blocks or exceedances above a large threshold, the proposed model uses all the data available efficiently, and hence not only fits the extremes but also models the entire rainfall distribution. We applied the method to both winter daily precipitation and weekly precipitation totals across a study region in south-west Western Australia. Our example shows that the proposed model can accommodate spatially and temporally varying volatility and skewness, and efficiently represents spatial and short-term temporal dependencies. \\
\indent In future research we plan to extend the method to more general treatments of similar spatio-temporal processes in different ways. Firstly the coefficients $\beta_k$ can be made spatially varying and temporally dynamic as in \cite{spdy1111}. Spatially varying coefficient models are often used to address the point-to-area problem of covariates which only vary over time, but whose impact may vary across space \citep{gelfand2003spatial}, whereas dynamic linear models \citep{west1985dynamic,stroud2001dynamic} are popular for modelling data with seasonal variations. Recent applications include \cite{dou2010modeling,ghosh2010spatio, mahmoudian2014spatio} and \cite{sahu2012comparison}. Secondly the autoregressive parameter $b$ can be generalised in a similar way \citep{spdy1111,cressie2011statistics}.   \\ 
\indent In this research, we adopted a censoring approach for modelling zero precipitation. Similiar approaches were also used in \cite{12qwe}, \cite{glasbey1997rainfall} and \cite{ex1}. An alternative approach for modelling zero precipitation considers a mixed probability distribution composed of a discrete at zero and a continuous distribution for non-zero precipitation; see e.g. \cite{srikanthan1999stochastic} and \cite{woolhiser1992modeling}. Another approach involves using the empirical distribution below a small threshold; see \cite{dupuis2012modeling}. It is also important to recognise that when analysing weekly precipitation totals, the data became less zero-inflated, and the censoring approach coupled with the flexibility of the generalised hyperbolic structure was able to model zero precipitation very well. Nevertheless, the proposed model could be adapted for the above alternative methods for modelling zero precipitation.   \\
\indent There are some limitations of the proposed method. First of all, the model will fail when applying to massive spatio-temporal dataset due to the big-n problem \citep{cressie2011statistics}. Second, in our research the temporal non-stationarity is modelled through heteroscedasticity but we assumed the spatial process is stationary, which might not be true for large study regions. However, complicated spatial and temporal interactions in the model make it difficult to generalise it to spatially non-stationary processes.
\restoregeometry
\newgeometry{top=1cm,left=1cm,right=1cm,bottom=1.75cm}
\begin{figure}[h]
 \centerline{\includegraphics[width=195mm]{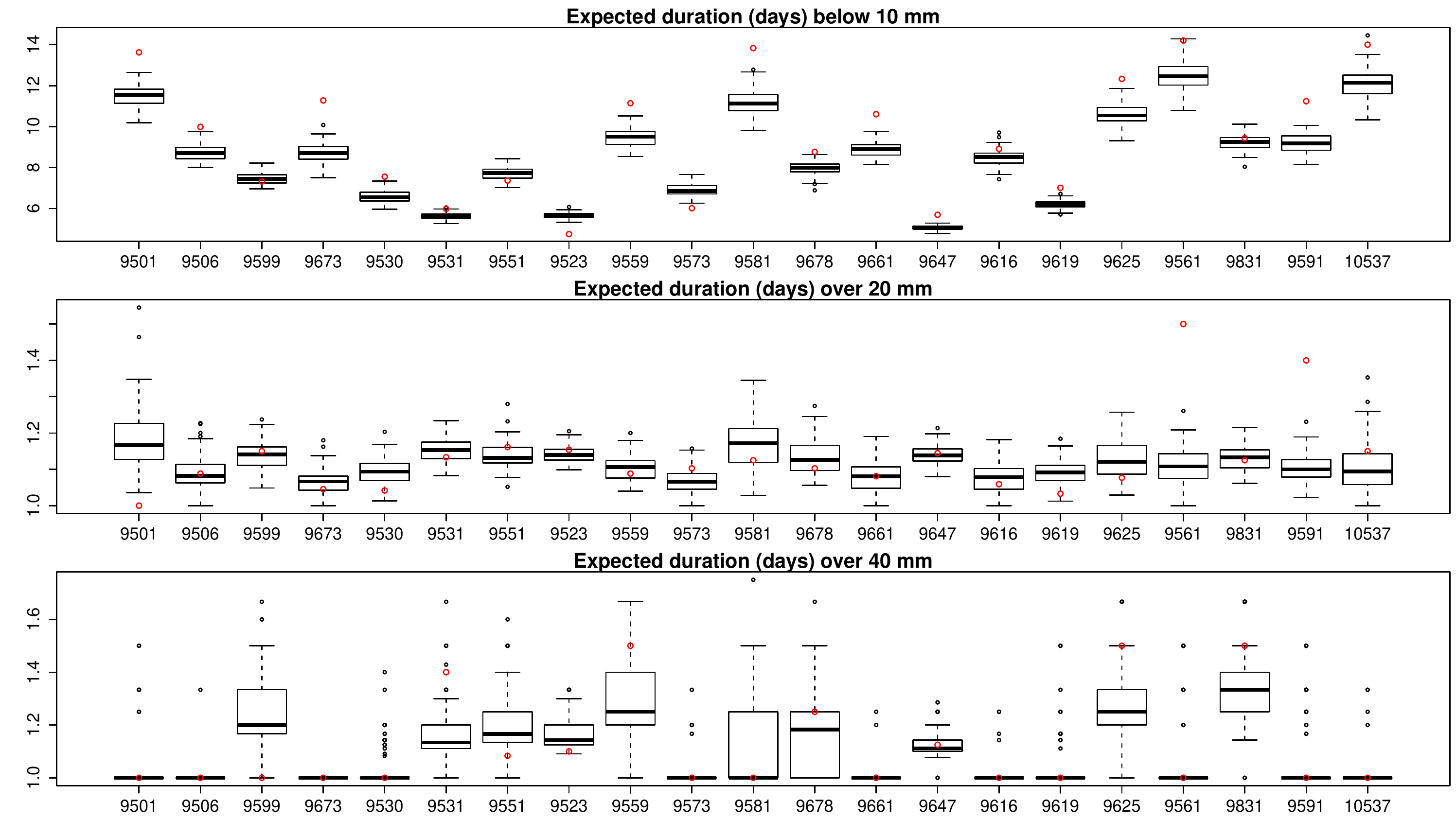}}
 \caption{\footnotesize Out-of-sample spatial predictions of expected durations (days) over/below thresholds based on the proposed hierarchical model and 1,000 simulations. The red dots are the mean duration over/below thresholds observed in year 1988-2008.}
\end{figure}
\begin{figure}[H]
 \centerline{\includegraphics[width=195mm]{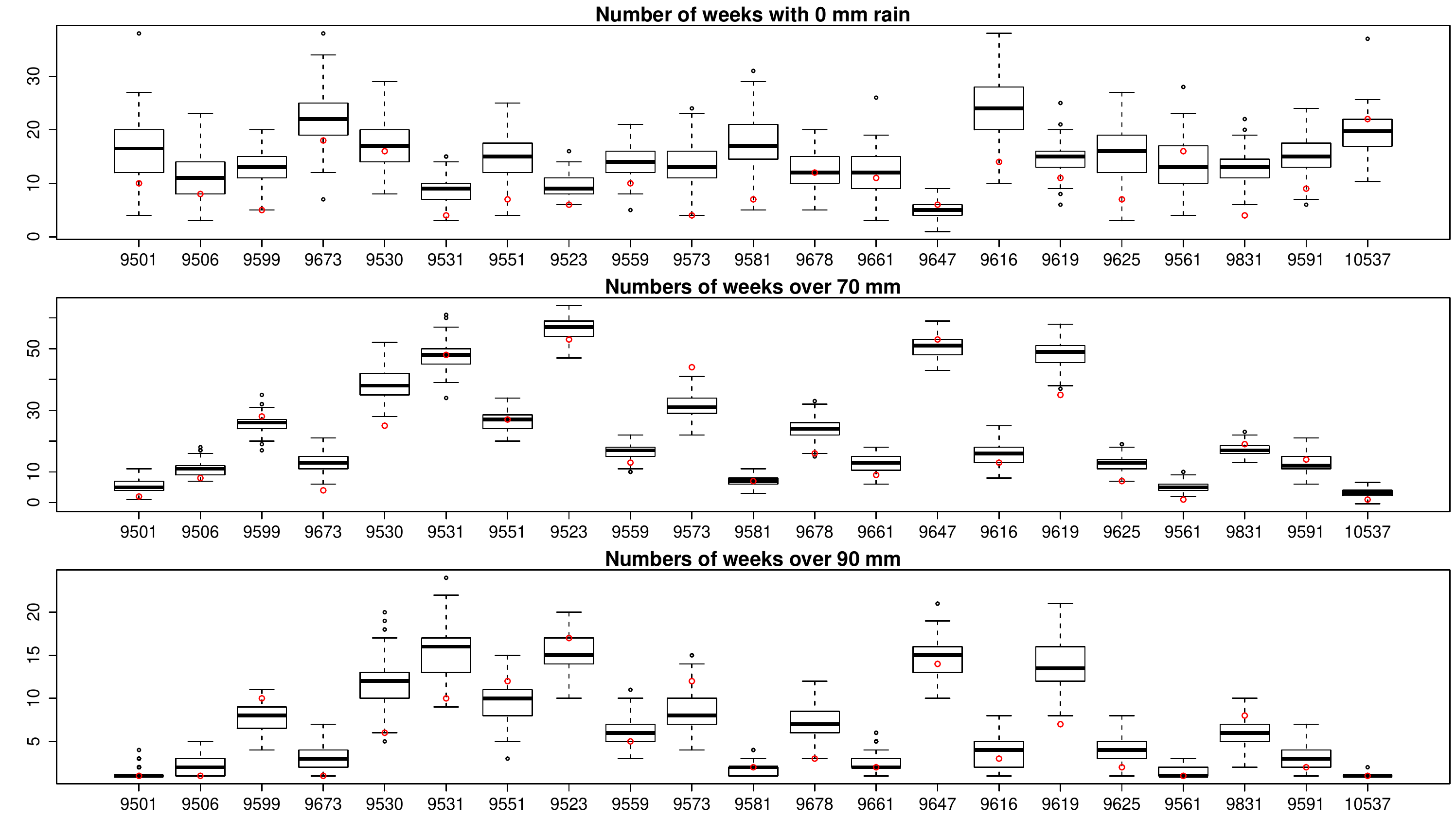}}
 \caption{\footnotesize Out-of-sample spatial predictions of numbers of weeks over/below thresholds based on the proposed hierarchical model and 1,000 simulations. The red dots are the numbers of weeks over/below thresholds observed in year 1958-2008.}
\end{figure}
\restoregeometry
\newpage
\newgeometry{left=0.1cm,right=0.1cm, bottom=2cm}

\begin{figure}[h]
        \centering
        \begin{subfigure}[h]{102mm}
                \includegraphics[width=109mm]{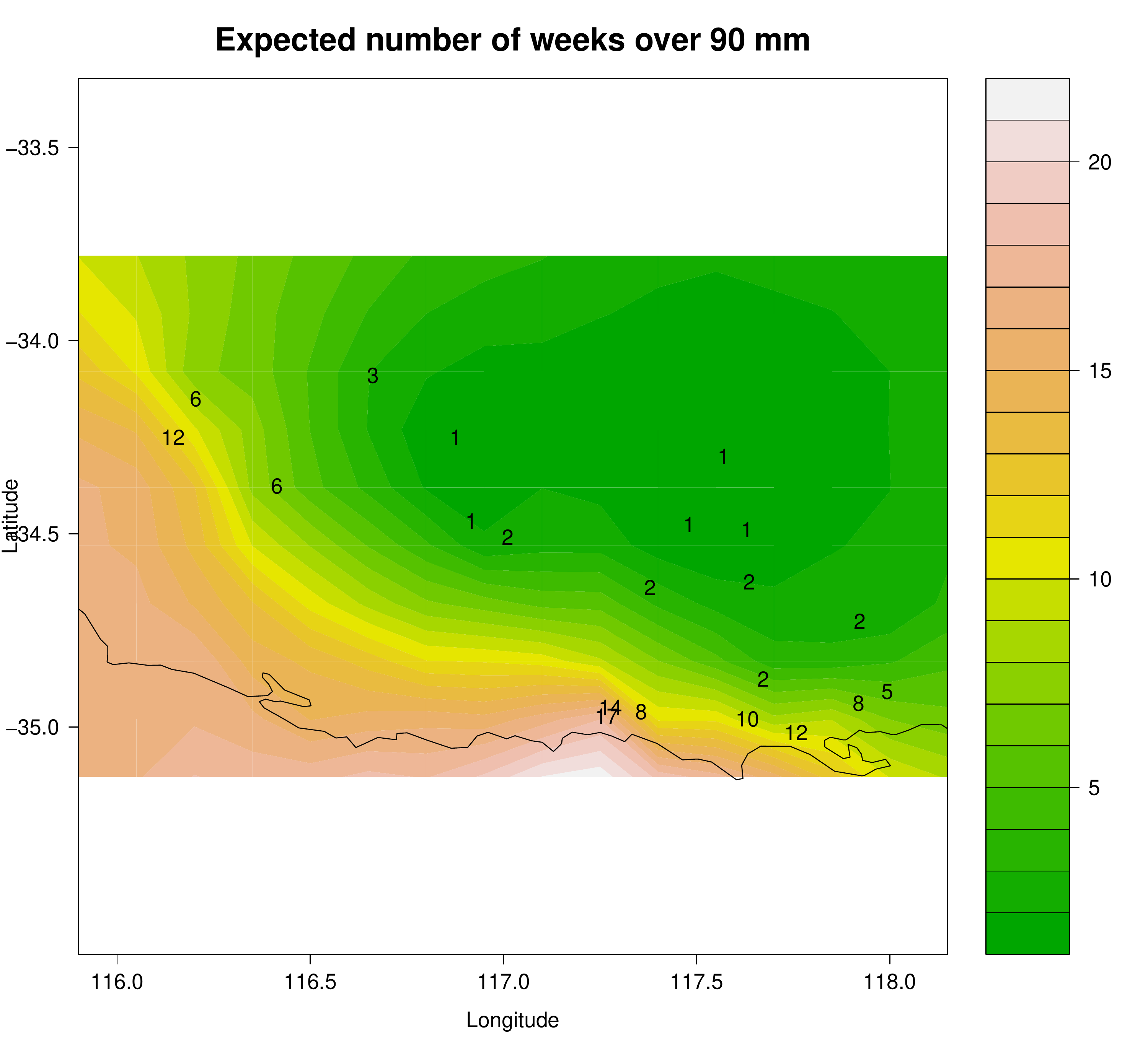}
        \end{subfigure}
     ~
        \begin{subfigure}[h]{100mm}
                \includegraphics[width=109mm]{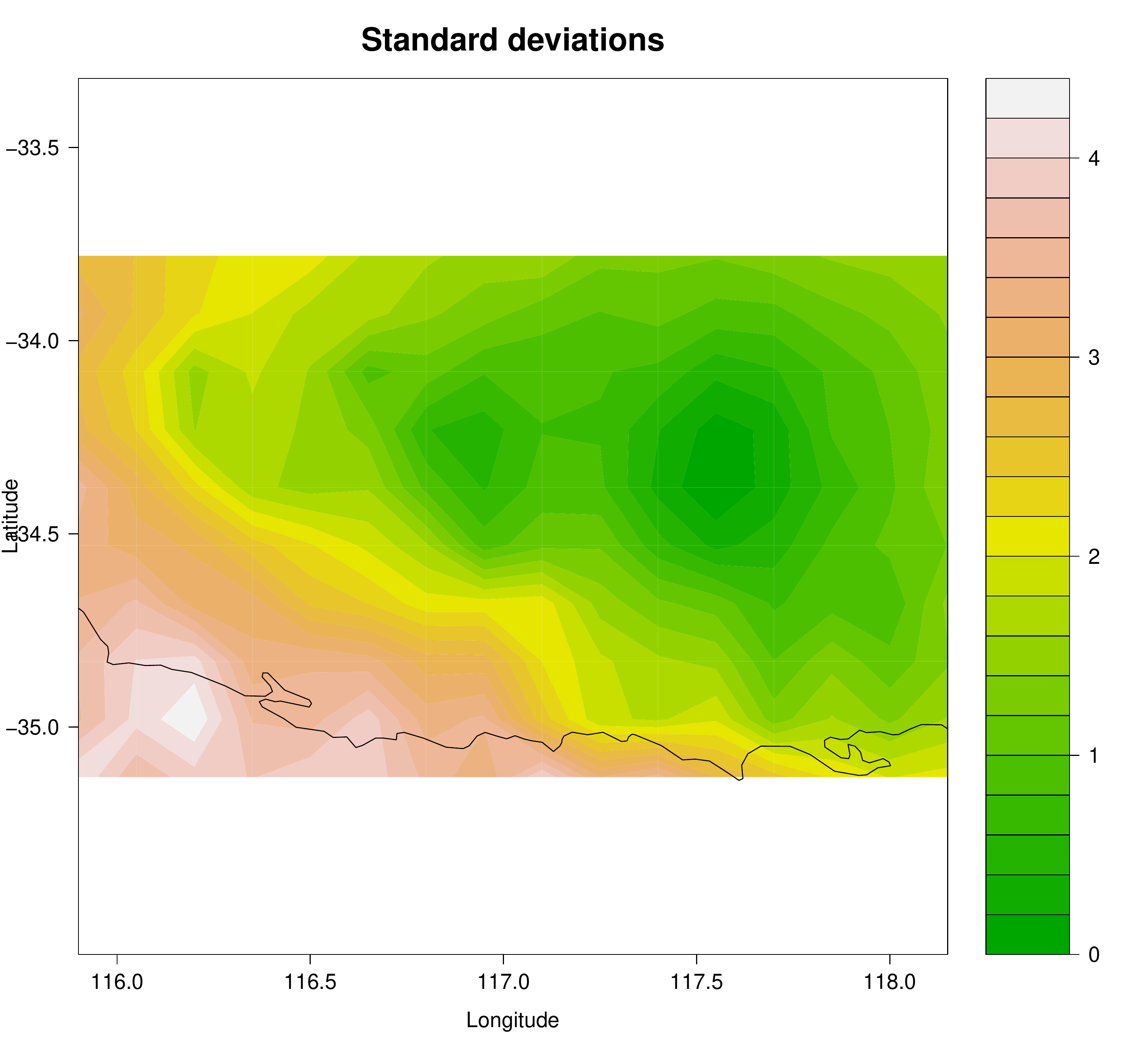}
        \end{subfigure}

        \begin{subfigure}[h]{102mm}
                \includegraphics[width=109mm]{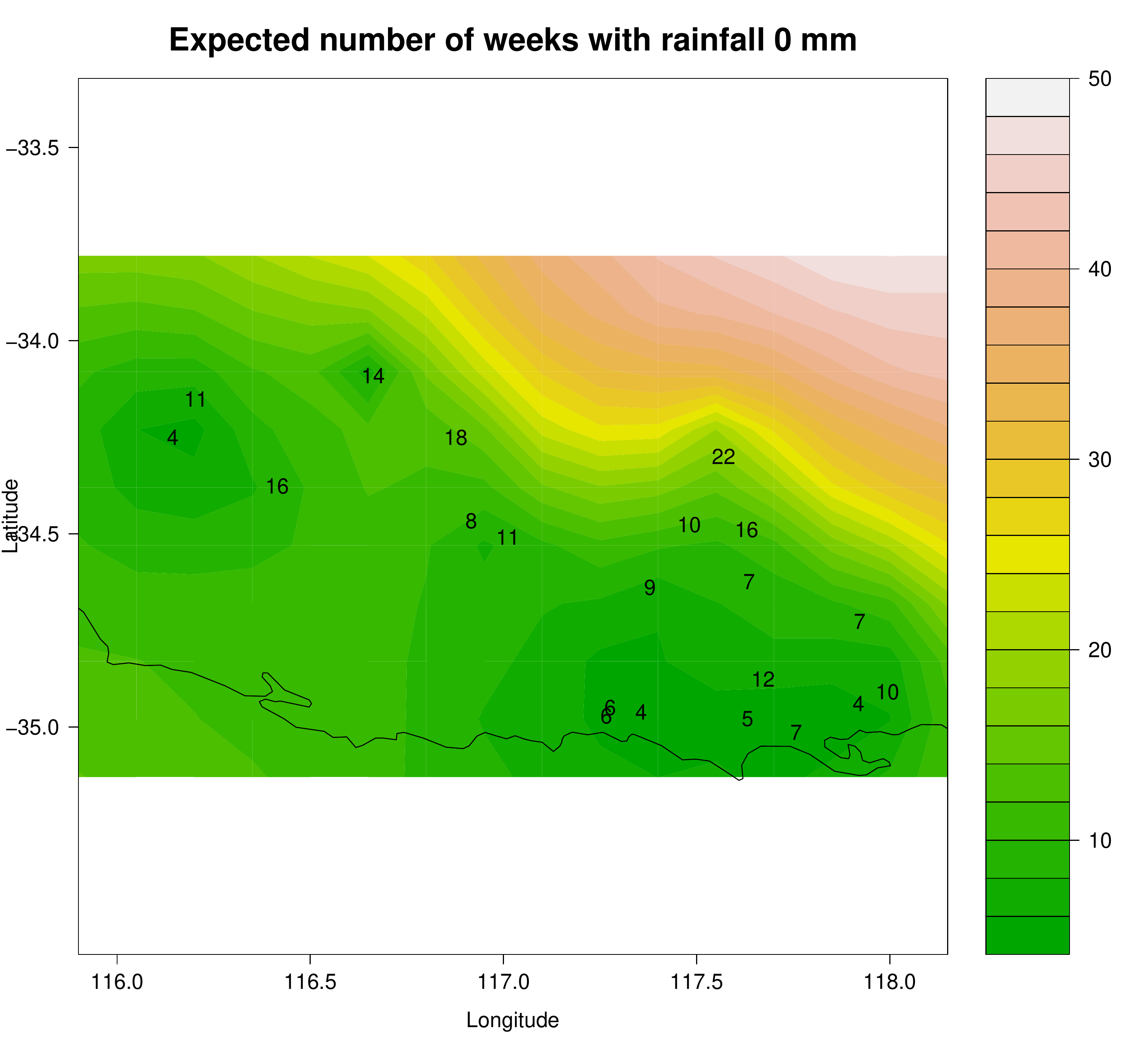}
        \end{subfigure}
     ~
        \begin{subfigure}[h]{100mm}
                \includegraphics[width=109mm]{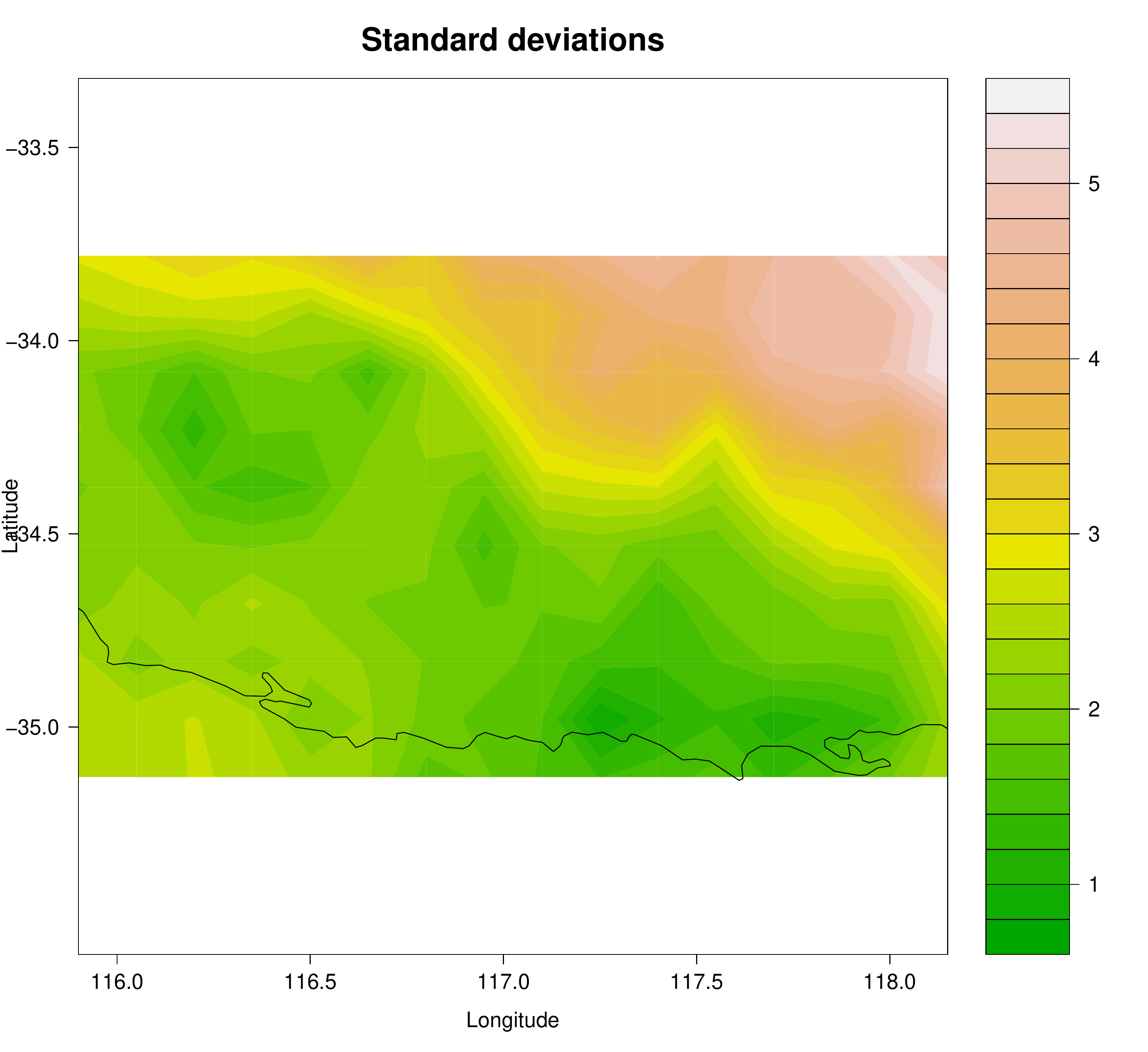}
        \end{subfigure}
 
        \caption{\footnotesize Predicted expected numbers of weeks with $>$90 mm and 0 mm rain and the corresponding standard deviations for the study region, based on the proposed hierarchical model and 1,000 simulations. The numbers represent the numbers of weeks with $>$90 mm and 0 mm rain observed in winter weekly precipitation sum in year 1958-2008.}
\end{figure}
\restoregeometry
\newpage
\newgeometry{bottom=2cm,top=1cm}
\section*{Appendix}
\begin{figure}[H]
 \centerline{\includegraphics[width=180mm]{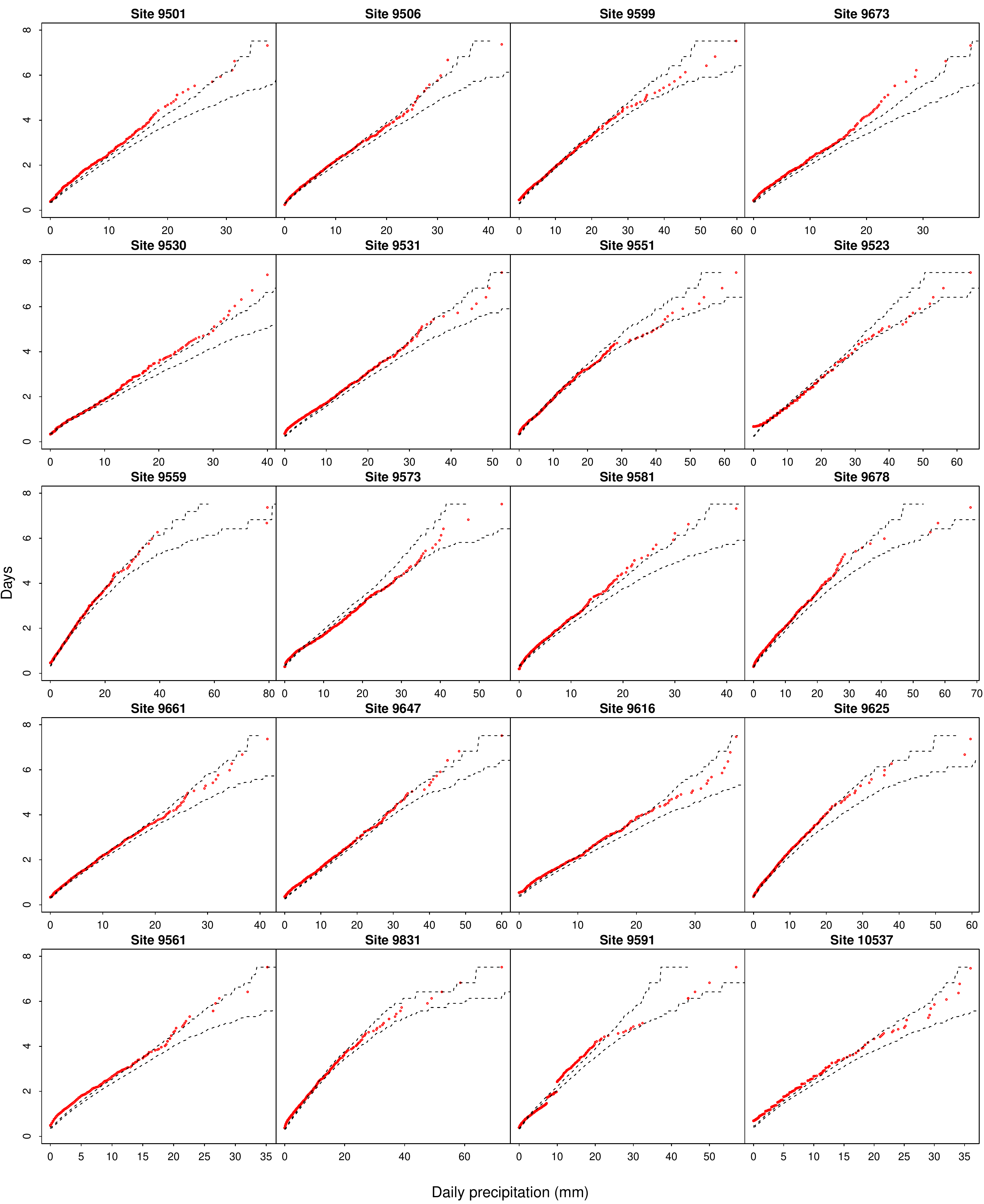}}
 \caption{\footnotesize Out-of-sample spatial predictions of precipitation return periods (days). The dashed lines are the 95\% CIs, which are based on the proposed hierarchical model and 1,000 simulations; the red dots are the observed precipitation return periods (days) of winter daily precipitation observed in year 1988-2008.}
\end{figure}
\restoregeometry
\begin{figure}[h]
 \centerline{\includegraphics[width=180mm]{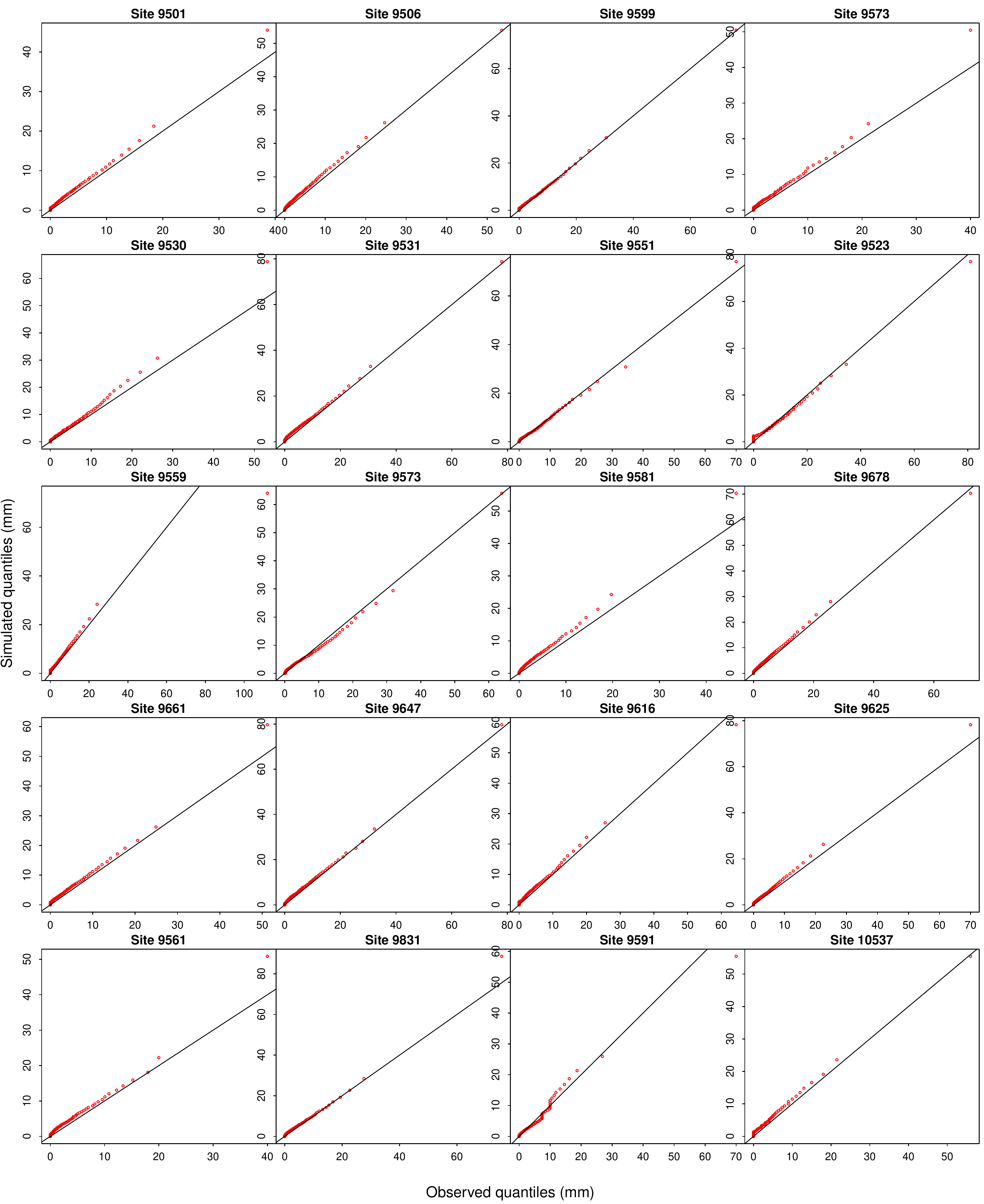}}
 \caption{\footnotesize Out-of-sample spatial predictions of distributions of daily precipitation based on the proposed hierarchical model and 1,000 simulations. Observed quantiles of winter daily precipitation observed in year 1988-2008 are on the x-axis, and mean simulated quantiles are on the y-axis.}
\end{figure}

\begin{figure}[h]
 \centerline{\includegraphics[width=180mm]{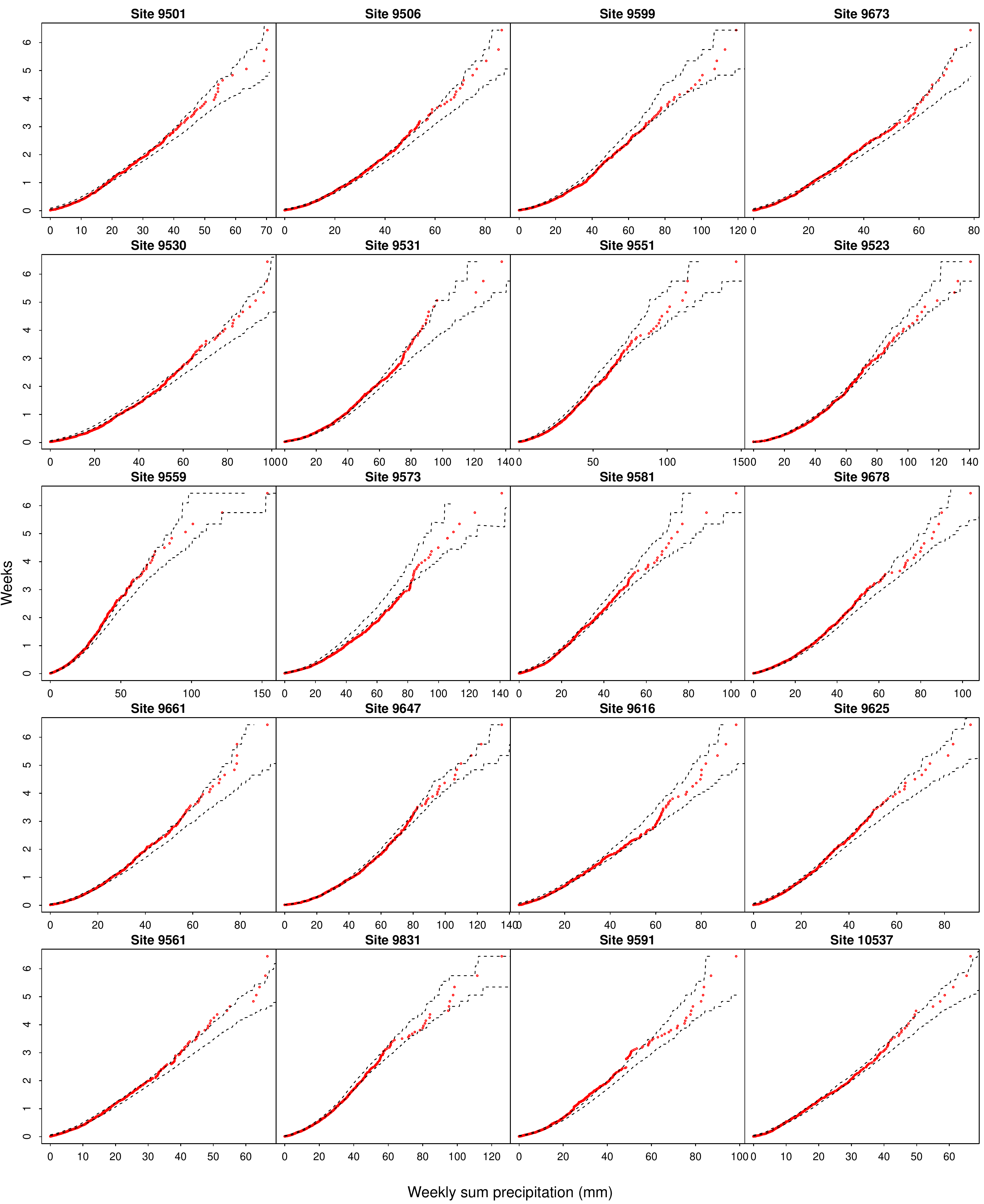}}
 \caption{\footnotesize Out-of-sample spatial predictions of precipitation return periods (weeks). The dashed lines are the 95\% CIs, which are based on the proposed hierarchical model and 1,000 simulations; the red dots are the observed precipitation return periods (days) of winter weekly precipitation sum observed in year 1958-2008.}
\end{figure}
\begin{figure}[h]
 \centerline{\includegraphics[width=180mm]{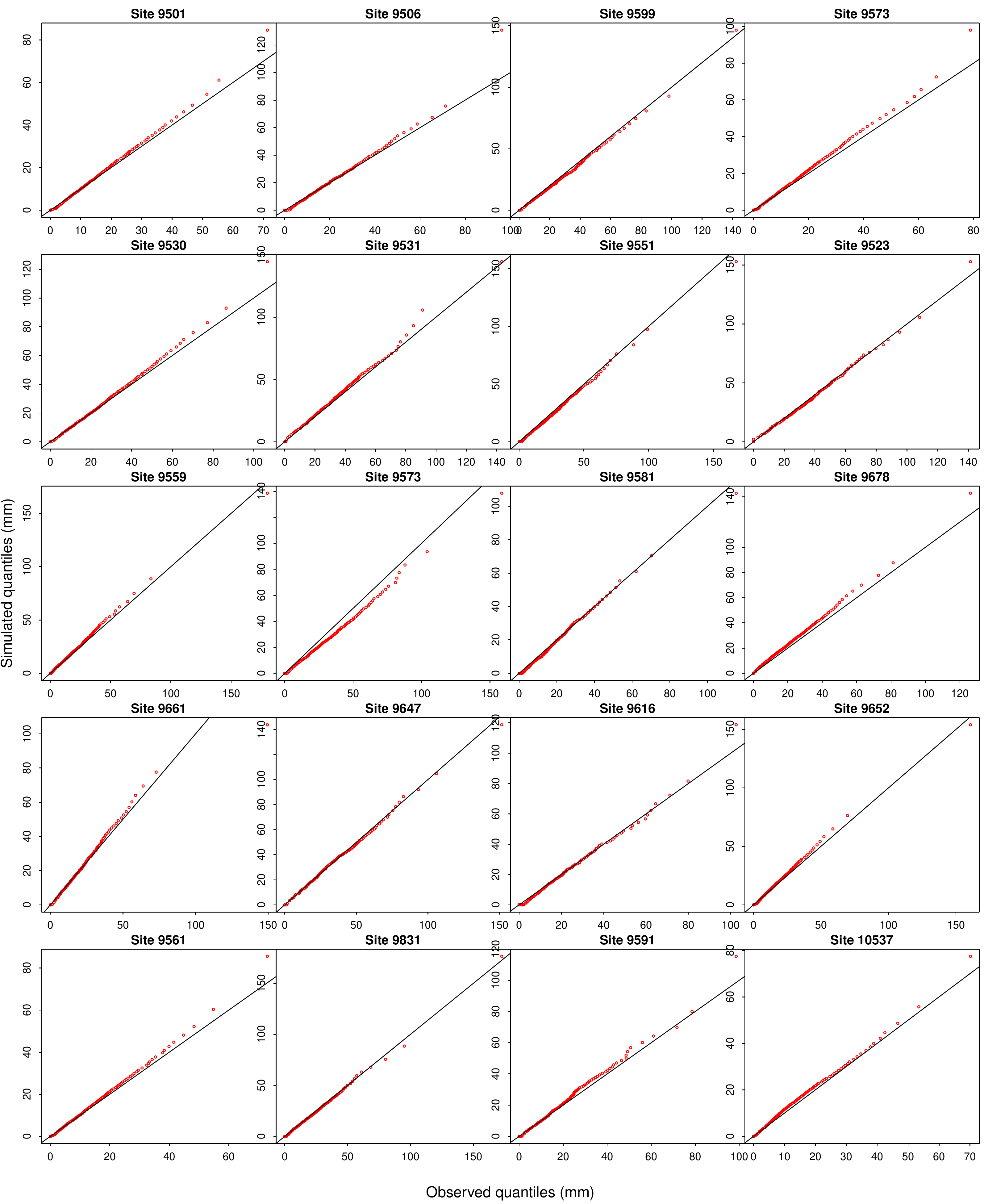}}
 \caption{\footnotesize Out-of-sample spatial predictions of distributions of weekly precipitation sum based on the proposed hierarchical model and 1,000 simulations. Observed quantiles of winter weekly  precipitation sum observed in year 1958-2008 are on the x-axis, and mean simulated quantiles are on the y-axis.}
\end{figure}

\clearpage
\bibliographystyle{Chicago}
\footnotesize
\newgeometry{left=1.85cm,right=1.85cm, bottom=2.25cm,top=1.75cm}
\bibliography{GH}

\end{document}